\documentclass[nofootinbib,aps,prl,twocolumn,showpacs,superscriptaddress, reprint]{revtex4-2}

\usepackage{CJK}
\usepackage[T1]{fontenc}
\usepackage[utf8]{inputenc}
\usepackage{graphicx}
\usepackage{natbib}
\usepackage{hyperref}
\usepackage{subfigure}
\usepackage[english]{babel}
\usepackage{array}
\usepackage{booktabs}
\usepackage{xcolor}
\usepackage{float}
\usepackage{amsmath}

\renewcommand{\thefootnote}{\fnsymbol{footnote}}

\usepackage{appendix}

\begin{document}
\title{Measurement of the cosmic \emph{p}+He energy spectrum from 50~GeV to 0.5~PeV with the DAMPE space mission}

\scriptsize
\noindent
\author{
 F.~Alemanno$^{1,2, \dag}$\footnotemark[2],
 C.~Altomare$^{3}$,
 Q.~An$^{4,5}$,
 P.~Azzarello$^{6}$, 
 F.~C.~T.~Barbato$^{1,2}$, 
 P.~Bernardini$^{7,8}$, 
 X.~J.~Bi$^{9, 10}$,
 I.~Cagnoli$^{1,2}$,
 M.~S.~Cai$^{11,12}$, 
 E.~Casilli$^{7,8}$
% ...
 \\
 (DAMPE Collaboration)*\footnotemark[1]
 \\
\noindent
$^1$Gran Sasso Science Institute (GSSI), Via Iacobucci 2, I-67100 L’Aquila, Italy \\
$^2$Istituto Nazionale di Fisica Nucleare (INFN) - Laboratori Nazionali del Gran Sasso, I-67100 Assergi, L’Aquila, Italy \\
$^{3}$Istituto Nazionale di Fisica Nucleare (INFN) - Sezione di Bari, I-70125, Bari, Italy \\
$^4$State Key Laboratory of Particle Detection and Electronics, University of Science and Technology of China, Hefei 230026, China \\
$^5$Department of Modern Physics, University of Science and Technology of China, Hefei 230026, China \\
$^6$Department of Nuclear and Particle Physics, University of Geneva, CH-1211 Geneva, Switzerland \\ 
$^7$Dipartimento di Matematica e Fisica E. De Giorgi, Università del Salento, I-73100, Lecce, Italy \\
$^8$Istituto Nazionale di Fisica Nucleare (INFN) - Sezione di Lecce, I-73100, Lecce, Italy \\
$^9$Institute of High Energy Physics, Chinese Academy of Sciences, Yuquan Road 19B, Beijing 100049, China \\
$^{10}$University of Chinese Academy of Sciences, Yuquan Road 19A, Beijing 100049, China \\
$^{11}$Key Laboratory of Dark Matter and Space Astronomy, Purple Mountain Observatory, Chinese Academy of Sciences, Nanjing 210023, China \\
$^{12}$School of Astronomy and Space Science, University of Science and Technology of China, Hefei 230026, China \\
$^{13}$Istituto Nazionale di Fisica Nucleare (INFN) - Sezione di Perugia, I-06123 Perugia, Italy \\
$^{14}$Institute of Modern Physics, Chinese Academy of Sciences, Nanchang Road 509, Lanzhou 730000, China \\ 
$^{15}$National Space Science Center, Chinese Academy of Sciences, Nanertiao 1, Zhongguancun, Haidian district, Beijing 100190, China \\
$^{16}$Dipartimento di Fisica “M. Merlin” dell’Università e del Politecnico di Bari, I-70126, Bari, Italy \\
$^{17}$Department of Physics and Laboratory for Space Research, the University of Hong Kong, Pok Fu Lam, Hong Kong SAR, China
}

\date{\today}

\begin{abstract}

Recent observations of the light component of the cosmic-ray spectrum have revealed unexpected features that motivate further and more precise measurements up to the highest energies. The Dark Matter Particle Explorer is a satellite-based cosmic-ray experiment that has been operational since December 2015, continuously collecting data on high-energy cosmic particles with very good statistics, energy resolution, and particle identification capabilities.   
In this work, the latest measurements of the energy spectrum of proton+helium in the energy range from 46~GeV to 464~TeV are presented. Among the most distinctive features of the spectrum, a spectral hardening at 600~GeV has been observed, along with a softening at 29~TeV measured with a 6.6$\sigma$ significance. Moreover, the detector features and the analysis approach allowed for the extension of the spectral measurement up to the sub-PeV region.  
Even if with small statistical significance due to the low number of events, data suggest a new spectral hardening at about 150~TeV.

\end{abstract}

\maketitle

\normalsize
\footnotetext[1]{dampe@pmo.ac.cn}

\renewcommand{\thefootnote}{\arabic{footnote}}

\emph{Introduction}. Cosmic rays (CRs) are high-energy particles accelerated in extreme regions of the Universe. CRs can be of galactic (GCRs) or extragalactic origin and they bring information about the astrophysical particle accelerators where they are generated and the interstellar or extragalactic medium they travel through. The cosmic-ray spectrum is composed of several nuclear species and it extends to energies beyond 10$^{20}$~eV. Shock acceleration mechanisms predict a single power-law energy spectrum for GCRs below the energy corresponding to the so-called ``all particle knee'' (3$-$4~PeV), which translates to an E$^{-2.6}-$E$^{-2.7}$ energy spectrum detected at Earth~\cite{Aloisio2018}. However, unexpected spectral features have been reported by several experiments for protons, helium, and heavier nuclei over the past years~\cite{Panov2009EnergySO, Yoon_2017,doi:10.1126/science.1199172,ADRIANI2013219,PhysRevLett.114.171103,PhysRevLett.115.211101,PhysRevLett.119.251101,PhysRevLett.120.021101,PhysRevLett.124.211102, doi:10.1126/sciadv.aax3793, PhysRevLett.122.181102, Ahn_2010,Atkin_2017, AHN20061950, DAMPECOLLABORATION20222162, CALET:2023nif, GREBENYUK20192546}. The spectrum of GCRs becomes harder around several hundreds of GeV and softer again above 10~TeV~\cite{doi:10.1126/sciadv.aax3793, PhysRevLett.126.201102,PhysRevLett.129.101102}. These deviations from a single power law motivate a deeper understanding of the CR acceleration and propagation mechanisms. Space-borne magnetic spectrometers like PAMELA and AMS have provided precise measurements of different CR species. However, they can only reach rigidities up to a few TV~\cite{doi:10.1126/science.1199172,ADRIANI2013219,PhysRevLett.114.171103,PhysRevLett.115.211101,PhysRevLett.119.251101,PhysRevLett.120.021101,PhysRevLett.124.211102}. Direct measurements of higher-energy CRs were performed with previous generation space and balloon-borne experiments, but with considerable statistical and instrumental uncertainties~\cite{AHN20061950,Atkin_2017,Ahn_2010,GREBENYUK20192546}.

The Dark Matter Particle Explorer (DAMPE) is a space-based particle and gamma-ray detector that has been operational since December 2015. It is designed to observe cosmic radiation up to $\sim$10~TeV for photons and e$^{-}$ + e$^{+}$, and hundreds of TeV for protons and ions while searching for indirect signatures of dark matter. The instrument consists of four subdetectors, the first being a plastic scintillator detector (PSD), designed to discriminate electrons from gamma rays and measure the absolute charge of impinging particles. The PSD comprises 82 bars, divided into 2 orthogonal layers, which are composed of 2 planes of staggered bars each. Below the PSD, a silicon-tungsten tracker-converter (STK) is used to measure the charged particle direction, giving additional information on the charge and converting photons in electron-positron pairs (with the help of tungsten layers). A bismuth germanium oxide (BGO) imaging calorimeter measures the energy of the particle and separates hadronic from electromagnetic showers. The BGO calorimeter is made of 14 layers, with 22 BGO bars each, for a total depth of more than 31 radiation lengths and $\sim$1.6 nuclear interaction lengths. Finally, the neutron detector, composed of boron-loaded plastic scintillators, collects neutrons from hadronic showers further refining the event identification. A schematic view of the DAMPE detector is shown in Fig. S1 of the Supplemental Material. DAMPE has a deep calorimeter, large acceptance, and good energy resolution ($\sim$1.5\% for electrons and $\sim$30\% for protons) making it an optimal instrument for measuring cosmic rays up to a few hundred TeV~\cite{CHANG20176}. In this study, the energy spectrum for \emph{p}+He is presented, using six years of flight data collected by DAMPE. By selecting a combined proton and helium sample, event selection criteria can be relaxed (with respect to the case of \emph{p} alone or He alone) while keeping low contamination and larger statistics. This allows for an extension in energy up to $\sim$0.5~PeV and provides for the first time a bridge between space-based and ground-based results with relatively small uncertainties.

\emph{Monte Carlo simulations}. Monte Carlo (MC) simulations are needed to understand the response of the detector to different particles. In this analysis, the GEANT4 version 4.10.5 toolkit~\cite{AGOSTINELLI2003250} is used along with the FTFP$\_$BERT physics list\footnote{\url{https://geant4.web.cern.ch/node/302}} for protons between 10~GeV and 100~TeV and helium nuclei between 10~GeV and 500~TeV. The physics list EPOS-LHC~\cite{PhysRevC.92.034906} is used for the energy interval 100~TeV$-$1~PeV for protons and 500~TeV$-$1~PeV for helium, by linking them to GEANT4 with the Cosmic Ray Monte Carlo package\footnote{\url{https://web.ikp.kit.edu/rulrich/crmc.html}}~\cite{Tykhonov:2019Tw}. Before launching DAMPE into space, several beam tests were performed at CERN, using ion beams of 40~GeV/n, 75~GeV/n, and 400~GeV/n~\cite{WEI2019177, ZHANG2020163139, Jiang_2020}. The data taken in the beam tests were compared with the simulations, showing a good agreement. The simulated events are initially generated with an isotropic spectrum, following an E$^{-1}$ dependence, and then reweighted during the analysis according to an E$^{-2.6}$ power law, following both theoretical expectations and experimental observations. As detailed later on, the exact shape of the energy spectrum used to weigh MC events negligibly affects our analysis results. Additional MC data are produced with alternative hadronic interaction models. Specifically, helium nuclei are simulated with FLUKA 2011.2x~\cite{BOHLEN2014211}, which uses the DPMJET3 model~\cite{Engel:1994vs, Roesler:2000he, Fedynitch:2015kcn}, while GEANT4-QGSP$\_$BERT is used for protons. The spectrum is computed anew using these MC samples, with the difference between the two spectra providing an estimate for the systematic uncertainty from the hadronic interaction model. 

\emph{Event selection}. In this study, 72 months of flight data taken between January 2016 and December 2021 are used. The events potentially affected by the South Atlantic Anomaly (SAA) region are excluded from the analysis. 
From this dataset, an event preselection is applied first, followed by a selection of \emph{p} or He particles. This procedure is applied both to MC and flight data. After subtracting the instrumental dead time, which is 3.0725 ms per event ($\sim$18\% of the operation time), the on-orbit calibration time ($\sim$1.6\%), a giant solar flare period between September 9 and September 13, 2017,\footnote{\url{https://solarflare.njit.edu/datasources.html}} and the SAA passage time ($\sim$5\%)~\cite{AMBROSI201918}, a total live time of $\sim$1.45$\times 10^{8}$~s remains, corresponding to $\sim$76\% of the total operation time.

\emph{(i) Preselection}. The preselection is based mainly on the measurements performed by the BGO calorimeter, according to the following criteria:

\begin{itemize}
    \item The energy deposited by a minimum ionizing particle (MIP) in a BGO bar is expected to be $\simeq$~23~MeV. The activation of the high-energy trigger (HET) is required, with the condition of an energy deposition larger than $\sim$ 10 MIPs in the first 3 BGO layers and larger than $\sim$ 2 MIPs in the fourth layer~\cite{Zhang_2019}. Events that are able to initiate a shower at the top of the calorimeter will satisfy this condition.
    \item Events with deposited energy in the calorimeter higher than 20~GeV are selected, to avoid the effect of the geomagnetic rigidity cutoff~\cite{2015EP&S...67...79T}.
    \item The energy deposited in any single layer of the BGO calorimeter has to be lower than 35\% of the total energy, in order to reject most of the events entering from the sides of the calorimeter.
    \item Additionally, a good lateral containment of the shower inside the calorimeter is achieved, by asking for the shower axis to be contained in a central region covering 93$\%$ of the calorimeter width. Furthermore, events whose maximum energy deposition occurs at the lateral edges of the calorimeter are rejected.
\end{itemize}

\emph{(ii) Track selection}. The track of the incoming particle is reconstructed by the STK~\cite{TYKHONOV201843}. In order to select the highest-quality events, the STK information is combined with measurements from other subdetectors. The first requirement is that the reconstructed track in the STK should match the shower axis in the BGO calorimeter. An additional requirement is that the STK track and the signal in the PSD are consistent. To achieve this, a PSD fiducial volume (covering $\sim$97$\%$ of the PSD active area, in the central region) is defined, with the condition of having the STK track projection within that specific volume.  

\begin{figure*}[htpb]
  \centering
  \subfigure{\label{sub:158}\includegraphics[width=0.3\textwidth]{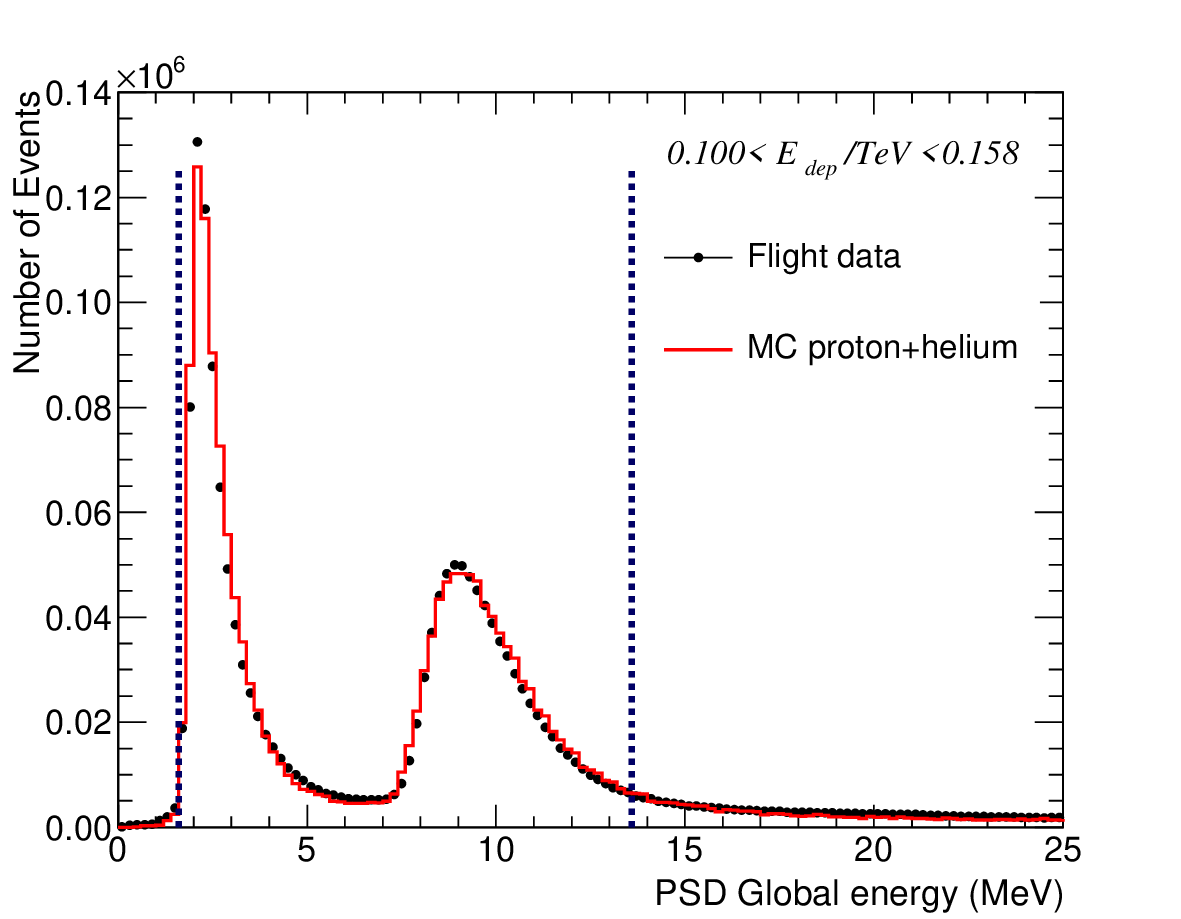}}%
  \,
  \subfigure{\label{sub:1585}\includegraphics[width=0.3\textwidth]{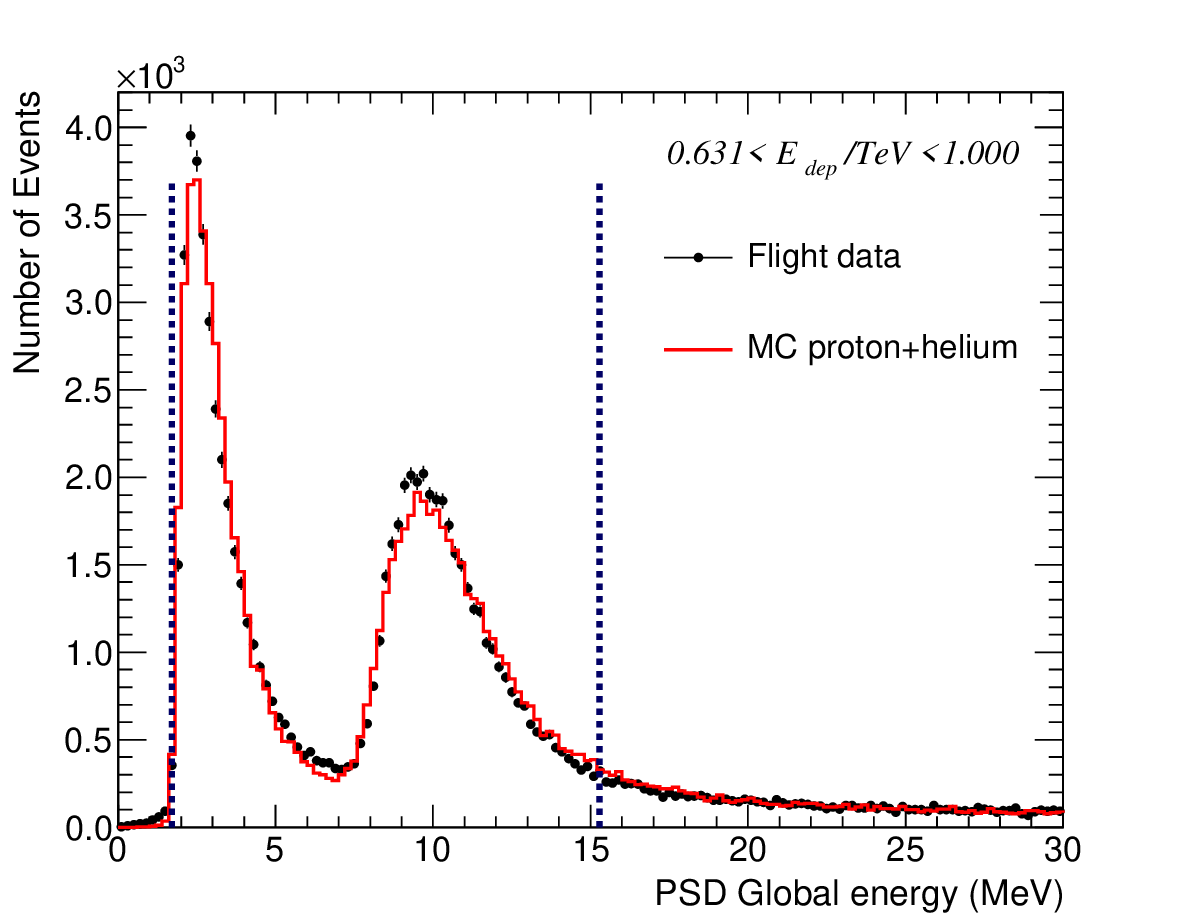}}%
  \,
  \subfigure{\label{sub:10000}\includegraphics[width=0.3\textwidth]{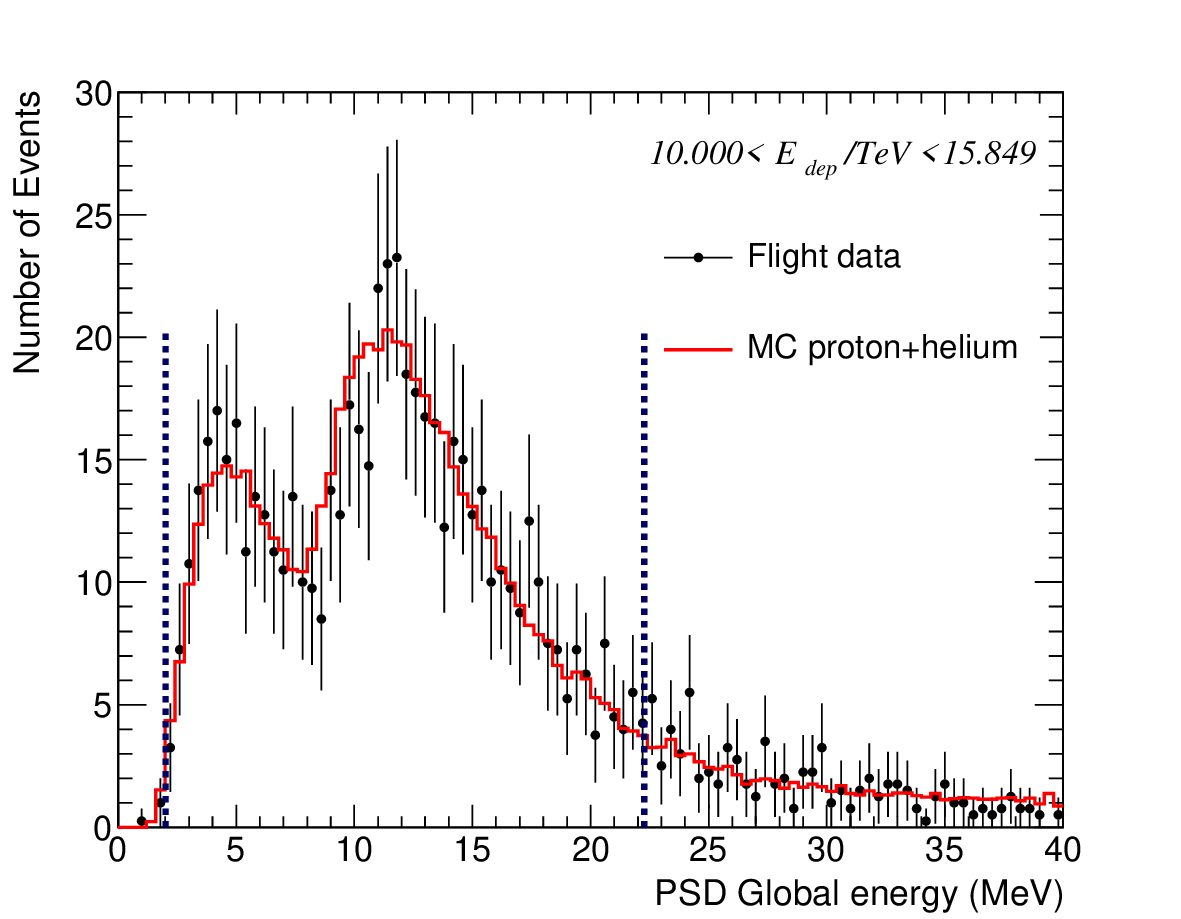}}%
  \caption{Distributions of PSD global energy, defined as the mean value of the energy released in the two PSD layers, for events with deposited energy in the BGO calorimeter in the ranges: 100–158~GeV (left), 0.631–1.00~TeV (center), and 10.0–15.8~TeV (right). Flight data are shown with black points, together with MC data of proton+helium, in red. The blue vertical dashed lines represent the charge selection ranges for \emph{p}+He.}
  \label{fig:PSD_dist}
\end{figure*}

\emph{(iii) Charge selection}. The different nuclei are selected according to the energy deposited in the PSD. A correction is applied to the signal of the PSD bars, accounting for light attenuation, detector alignment, and incident angle~\cite{DONG201931, Ma_2019}. After this correction, the signal can be considered to be proportional to $\rm Z^2$, in accordance with the Bethe-Bloch equation (with \emph{Z} being the charge of the incident particle). The PSD global deposited energy for a particular event is obtained by combining the independent energy loss readings from each of the 2 PSD layers. The deposited energy loss not only depends on the charge of the incident particle but also on its primary energy. For this reason, the charge selection is performed in different bins of energy deposited in the BGO calorimeter. For each bin, the PSD global energy distribution of the events is fitted using a Landau convoluted with a Gaussian function (LanGaus). The Landau function describes fluctuations in energy loss of ionizing particles and the Gaussian is used to account for detector effects. From these fits, the most probable value (MPV) and width value (sigma) of the resulting function are obtained. Both the MPV and sigma values have a dependence on the total energy deposited in the calorimeter. This dependence is modeled by fitting the MPV and sigma obtained from the fits with a fourth-order polynomial function of the logarithm of the energy, which is used to retrieve a charge selection condition for different values of deposited energy. The functions obtained for flight data and MC data were found to have a slight disagreement, probably because of an overestimation of the back-scattering effect in MC simulations. In order to account for this mismatch, a smearing correction is applied to the charge distributions for MC results: the proton and helium peaks are corrected in order to match MPV and sigma of flight data. Figure~\ref{fig:PSD_dist} shows an example of the PSD charge distributions for three different bins of deposited energy and their comparison with MC data, after the smearing correction. The vertical dashed lines show the charge selection conditions, with a maximum value of MPV$_{\rm He}$ + 6$\sigma_{\rm He}$ and a minimum value of MPV$_{\rm p}$ - 2$\sigma_{\rm p}$, where the sigma value is given by $\sigma=\sqrt{Width_{\rm Landau}^{2}+\sigma_{\rm Gaus}^{2}}$. These limits are optimized and chosen to maximize the statistics while maintaining a low background level ($\lesssim$ 0.4\% up to 10~TeV, see background section).
Finally, it was checked that using the STK as charge detector, and therefore adding events that do not cross the PSD, does not significantly increase the acceptance (less than 5\% and only at low energies) given the requirements of the lateral shower containment in the calorimeter.

\begin{figure}[htbp]
     \centering
     \includegraphics[width=0.45\textwidth]{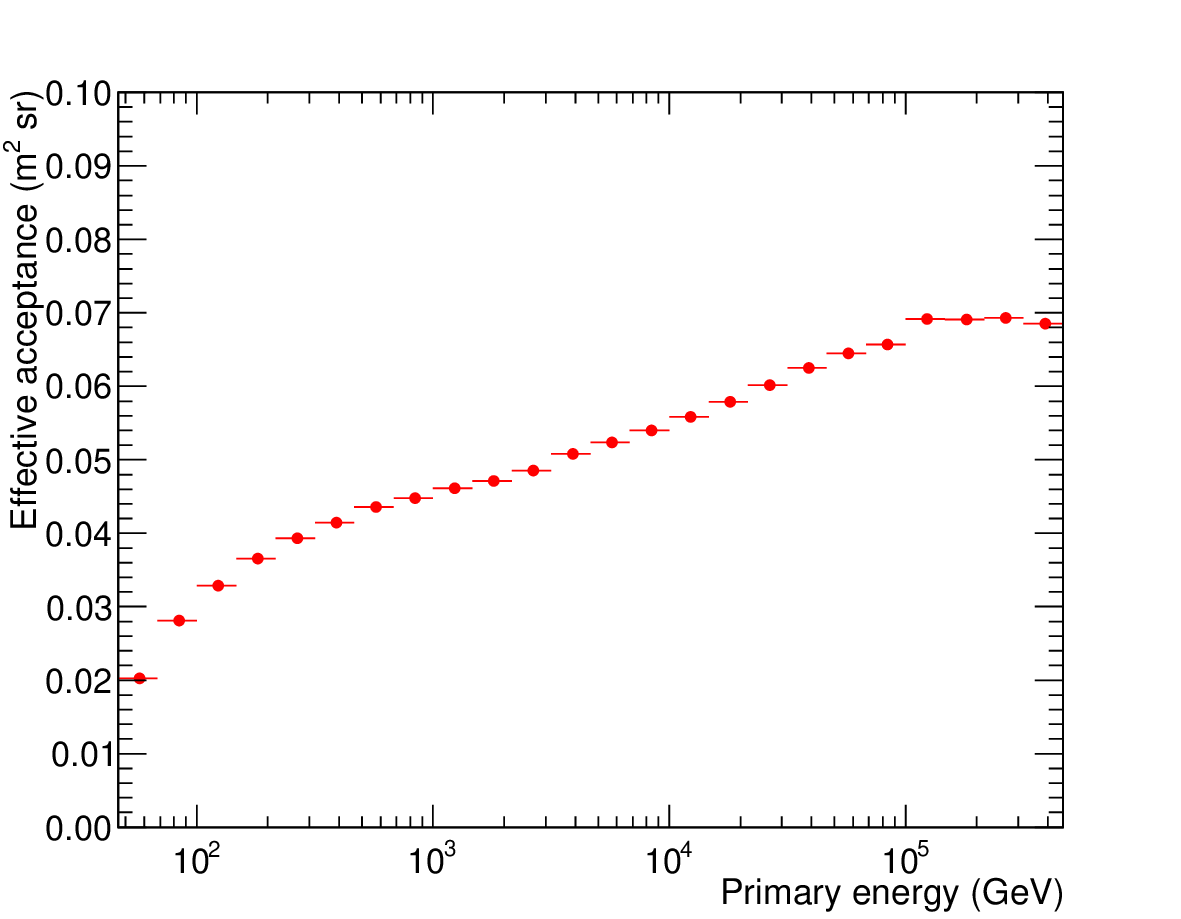}
     \caption{Effective acceptance of the \emph{p}+He analysis obtained by using \emph{p} and He MC samples, after applying all the selection cuts (see text).}
     \label{fig:acceptance}
 \end{figure}

\emph{Effective acceptance}. After applying the selection cuts described in the previous section, the efficiencies are computed using MC simulations. Their comparison with flight data and subsequent validation is reported in the Supplemental Material. Afterwards, the effective acceptance (${A^{\rm i}}$) can be evaluated. Figure~\ref{fig:acceptance}  shows the acceptance of the DAMPE detector as a function of the primary energy, which can be described by the following expression:
\begin{equation}
{A^{\rm i}(E_{\rm T}^{\rm i}) = G_{\rm gen}\times \frac{N(E_{\rm T}^{\rm i},\rm sel)}{N(E_{\rm T}^{\rm i})}},
\label{Formula:Acceptance}
\end{equation}
where $G_{\rm gen}$ is the geometrical factor used to generate MC data, $N(E_{\rm T}^{\rm i})$ is the number of MC generated events in the \emph{i}th bin of primary energy ($E_{\rm T}$), and $N(E_{\rm T}^{\rm i},\rm sel)$ the number of those MC events surviving the selection cuts. This result was found to be independent of the spectral shape or the \emph{p}/He mixture assumed in the simulation (see the section on the systematic uncertainty evaluation).

\emph{Background estimation}. Protons constitute a background for helium, and vice versa. By combining these nuclei in a single spectrum, the remaining background is very low and mainly comprises electrons-positrons and lithium nuclei. Electrons and positrons are separated from protons in the BGO calorimeter using shower morphology discrimination. A detailed description of the separation of electrons and positrons from protons can be found in~\cite{electrons}. For the present analysis, the contamination of electrons in the \emph{p}+He spectrum is $\sim$0.5$\%$ at 40~GeV of energy deposited in the BGO calorimeter, and it decreases with increasing energy. The lithium background is estimated using the template fit of the energy released in the PSD based on MC simulations of proton, helium, and lithium. The contamination from lithium is lower than 0.3$\%$ up to 10~TeV, and it is $\sim$1.6$\%$ for energies higher than 10~TeV. The background from electrons-positrons and lithium is shown in Fig.~S5 of the Supplemental Material.

\emph{Energy measurement and unfolding procedure}. The energy of the hadronic showers cannot be completely contained in the calorimeter. In particular, for \emph{p} and He, around 35\% to 45\% of the total energy is collected in the detector. Consequently, an unfolding procedure is necessary to obtain the energy spectrum of the incident particles. In this case, a Bayesian approach is adopted~\cite{DAGOSTINI1995487}, in which the detector response is estimated from MC simulations of both proton and helium nuclei, after applying the selection cuts described in the \emph{Event Selection} section. The actual number of events in the \emph{i}th bin of true energy, $N(E^{\rm i}_{\rm T})$, can be obtained from the following expression:
\begin{equation}
    N(E^{\rm i}_{\rm T}) = \sum_{j=1}^{n}P\left(E^{\rm i}_{\rm T} | E^{\rm j}_{\rm O}\right) N(E^{\rm j}_{\rm O}),
\end{equation}
where $N(E^{\rm j}_{\rm O})$ is the number of observed events in the \emph{j}th bin of energy deposited in the calorimeter ($E^{\rm j}_{\rm O}$) and $P\left(E^{\rm i}_{\rm T} | E^{\rm j}_{\rm O}\right)$ the response matrix derived from MC simulations (see Fig.~S6 of the Supplemental Material). 
The energy of an event is determined from the BGO calorimeter measurements, which needs to be corrected in order to obtain the true energy deposited in the calorimeter. For events with deposited energy $\gtrsim$ 4~TeV in a single BGO bar, some readout channels might get saturated. For this reason, a method developed using MC simulations is used to correct saturated events~\cite{YUE2020164645}. Another correction is applied to account for Birk's quenching in the BGO calorimeter. Quenching is more significant for heavy nuclei which produce more secondary particles with high charge and low velocity~\cite{Birks_1951}. The BGO quenching is taken into account by including its effect in the MC simulations for ionization energy densities above 10~MeV/mm~\cite{Quenching_BGO}. The effect is more important for incident energies below $\sim$80~GeV, where it would result in a $\sim$2\% lower energy reconstruction.

\begin{figure*}[htbp]
  \centering
  \subfigure{\label{sub:Direct}\includegraphics[width=0.45\textwidth]{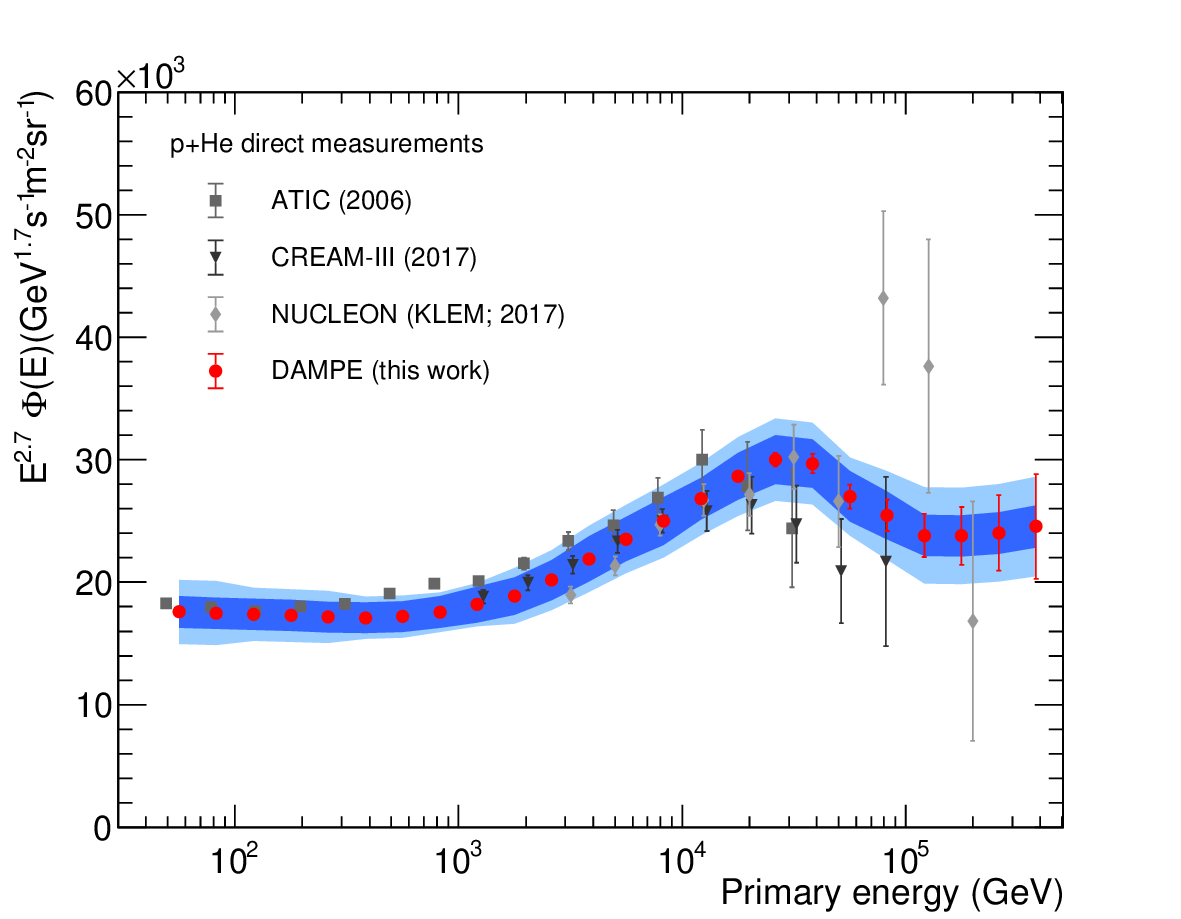} }%
  \,
  \subfigure{\label{sub:Indirect}\includegraphics[width=0.45\textwidth]{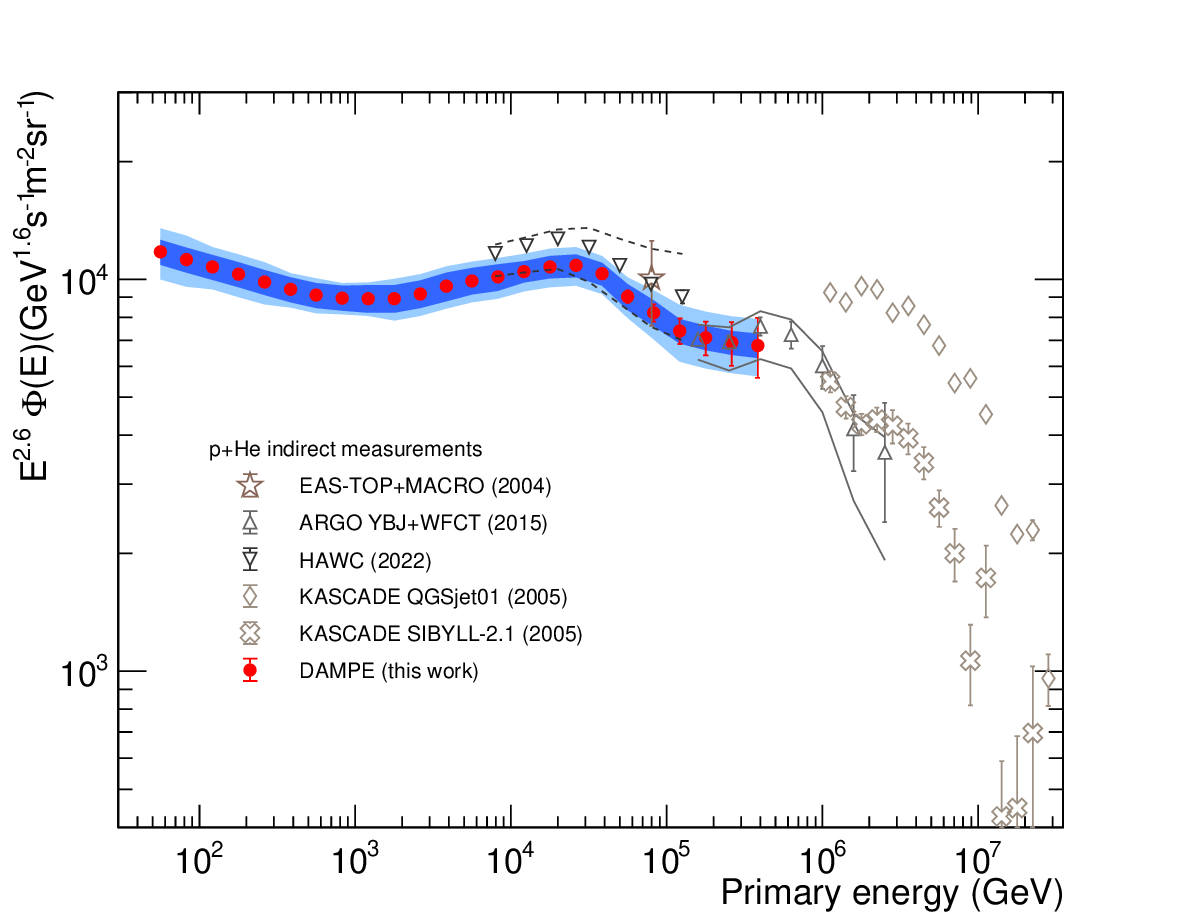}}%
  \caption{\emph{p}+He spectrum measured with the DAMPE detector (red circles), between 46~GeV and 464~TeV, compared with: direct measurements of \emph{p}+He made by ATIC-02~\cite{AHN20061950}, NUCLEON~\cite{Atkin_2017} and CREAM~\cite{Ahn_2010} (left), and indirect measurements from ARGO-YBJ+WFCT~\cite{PhysRevD.92.092005}, HAWC~\cite{PhysRevD.105.063021}, KASCADE~\cite{ANTONI20051} and EAS-TOP+MACRO~\cite{AGLIETTA2004223} (right). Statistical uncertainties (1$\sigma$) are represented by error bars, while the continuous bands represent the systematic uncertainties on the analysis (inner band) and the total systematic uncertainties (outer band). \label{exp_comparison}}%
\end{figure*}

\emph{Results}. The flux for each energy bin $\left(\Phi_{\rm i}\right)$ can be written as follows:
\begin{equation}
    \Phi_{\rm i}=\frac{\Delta N_{\rm i}}{\Delta T \times A_{\rm i} \times \Delta E_{\rm i}},
\end{equation}
with $N_{\rm i}$ the number of events in the \emph{i}th energy bin after the unfolding, $\Delta T$ the total live time, $A_{\rm i}$ the acceptance in the \emph{i}th bin, and $\Delta E_{\rm i}$ representing the width of the \emph{i}th energy interval. Figure~\ref{exp_comparison} shows the \emph{p}+He flux in the energy range 46~GeV$-$464~TeV, multiplied by a power of the energy and compared with other direct (Fig.~\ref{exp_comparison}, left) and indirect (Fig.~\ref{exp_comparison}, right) \emph{p}+He measurements. The 1$\sigma$ statistical uncertainties on DAMPE data are represented by error bars, while the continuous bands indicate the systematic uncertainties associated with the analysis procedure (inner band) and the total systematic uncertainties (outer band), including the one on the hadronic interaction model. The results are also reported in Table~S1 of the Supplemental Material.

The statistical uncertainties are associated with the Poissonian fluctuations of the number of detected events and the MC sample. However, due to the unfolding process, this uncertainty cannot be directly translated into the incident energy bins. To achieve this, a batch of toy-MC samples is generated according to a Poisson distribution for each deposited energy bin. The fluxes are then obtained through the regular unfolding procedure,  and their root mean square in each energy bin is taken as the 1$\sigma$ statistical error~\cite{doi:10.1126/sciadv.aax3793}.
\\The systematic uncertainty band is the result of several contributions. The main contribution (up to $\sim$15\%, for energy larger than 100~TeV) comes from the hadronic interaction model used for the MC simulation. The GEANT4-FTFP$\_$BERT model is found to be in better agreement with flight and test beam data~\cite{Vitillo, WEI2019177, ZHANG2020163139}, and is therefore chosen for the computation of the \emph{p}+He spectrum. To quantify the uncertainty resulting from this choice, the \emph{p}+He spectrum is computed also using the FLUKA DPMJET-3 model for helium and GEANT4-QGSP$\_$BERT for protons. The difference between the two spectra is used to estimate the uncertainty on the hadronic model. Additional contributions to the systematic uncertainties are given by the event selection procedure. In this case, selection efficiencies of MC and flight data are compared (more details are given in the Supplemental Material). Their difference is found to be $\sim$4\% for the HET efficiency, $\sim$2\% for the track selection efficiency, and a maximum of $\sim$3\% for the charge selection efficiency at energies higher than 2~TeV, as shown in Figs.~S2, S3, and S4 of the Supplemental Material. The quadratic sum of the aforementioned differences in efficiencies (between MC and data) is taken as the systematic uncertainty of the acceptance, amounting to $\sim$5.4 \% for energies higher than 2~TeV. A conservative approach is used and these uncertainties are regarded as symmetric. Another source of uncertainty is the assumed proton and helium mixture in the simulation. 
Several mixtures have been tested to evaluate the possible systematic effect on the final flux, including the ones coming from the DAMPE~\cite{doi:10.1126/sciadv.aax3793, PhysRevLett.126.201102} or AMS-02 \emph{p} and He measurements~\cite{PhysRevLett.114.171103,PhysRevLett.115.211101}. The result is a variation in the final flux always limited to 2$-$5\%. It was therefore decided to use the 50\% relative abundance as default value, so as not to introduce any model dependence or bias (e.g., due to the extrapolation to high energies of proton and helium measured fluxes), and the above 2$-$5\% was added as a symmetric contribution to the systematic uncertainties. For more information on the systematic uncertainties and their energy dependence please refer to Sec.~IV and Fig.~S7 of the Supplemental Material.  

The proton+helium spectrum has been fitted with a smoothly-broken power-law (SBPL) function following a similar approach to the one used in~\cite{PhysRevD.95.082007,electrons, doi:10.1126/sciadv.aax3793, PhysRevLett.126.201102} (see details in the Supplemental Material). The result shows the presence of a spectral hardening around $\sim$ 600~GeV followed by a softening at 28.8 $\pm$ 4.5~TeV, measured with a significance of 6.6$\sigma$. The hardening feature is in line with results obtained by other experiments~\cite{doi:10.1126/science.1199172, ADRIANI2013219, PhysRevLett.114.171103, PhysRevLett.115.211101, PhysRevLett.119.251101, PhysRevLett.120.021101, PhysRevLett.124.211102, PhysRevLett.126.041104, PhysRevLett.122.181102, Panov2009EnergySO, Ahn_2010, Yoon_2017, Atkin_2017} and previous results from DAMPE on proton~\cite{doi:10.1126/sciadv.aax3793} and helium~\cite{PhysRevLett.126.201102} spectra. Moreover, DAMPE revealed a softening feature in both the proton and helium spectra~\cite{doi:10.1126/sciadv.aax3793,PhysRevLett.126.201102}, further confirmed by the present analysis (the fit is shown in Fig.~S9 of the Supplemental Material). The spectral fit parameters obtained with the three analyses are reported in Table \ref{Table:SBPL}, where $E_{\rm b}$ is the energy in which there is a change of slope in the spectrum, $\gamma$ represents the spectral index before $E_{\rm b}$ and $\Delta\gamma$ the difference between the 2 indexes, before and after $E_{\rm b}$. The $E_{\rm b}$ values suggest rigidity-dependent features, even though a mass dependence cannot be ruled out. 

\begin{table}[htbp]
\centering
\caption{Results of the SBPL fit in the softening energy region for the DAMPE proton~\cite{doi:10.1126/sciadv.aax3793}, helium~\cite{PhysRevLett.126.201102} and \emph{p}+He spectra (this work). For the helium and \emph{p}+He results, the systematic uncertainties from the hadronic model are represented by the second uncertainty.}
\setlength{\extrarowheight}{2pt}
\begin{tabular}{ c  c  c  c}
\toprule
  & Proton & Helium & Proton+Helium \\ 
\midrule
 $E_{b}$ (TeV) & $13.6^{+4.1}_{-4.8}$ & $34.4^{+6.7+11.6}_{-9.8-0.0}$& $28.8_{-4.4-0.0}^{+6.2+2.9}$\\  
 $\gamma$ & $2.60\pm 0.01$ & $2.41^{+0.02+0.02}_{-0.02-0.00}$& $2.512_{-0.024-0.00}^{+0.021+0.01}$\\
 $\Delta\gamma$ & $-0.25 \pm 0.07$ & $-0.51^{+0.18+0.01}_{-0.20-0.00}$& $-0.427^{+0.057+0.00}_{-0.066-0.066}$ \\
\bottomrule
\end{tabular}
\label{Table:SBPL}
\end{table}

The DAMPE \emph{p}+He spectrum is in agreement with other direct-detection experiments, within the systematic uncertainties, and also shows good compatibility with the sum of the individual DAMPE proton and helium spectra (Fig. S8 of the Supplemental Material). The DAMPE \emph{p}+He spectrum suggests a second hardening at about 150~TeV, even if the low event statistics do not allow for a firm conclusion. Above 30~TeV the spectrum is well fitted with a broken power law, with $E_b$=110~$\pm$~53~TeV and spectral indexes of $\gamma_1$=2.91~$\pm$~0.07 and $\gamma_2$=2.68~$\pm$~0.14, before and after the break, respectively. The hypothesis of a single power law, above 30~TeV, with spectral index equal to $\gamma_1$ is disfavored at the level of 1.5$\sigma$. 
An evidence for a spectral hardening at about 166 TeV has been recently reported, after this work was submitted, by the GRAPES-3 ground-based experiment~\cite{PhysRevLett.132.051002}.
Moreover, hints for the same hardening have been previously given by HAWC (\emph{p}, He, and \emph{p}+He) and ISS-CREAM (\emph{p} only)~\cite{Abeysekara:2021ct, Choi_2022}, and foreseen in~\cite{Yue_2019}.
The above-described spectral features (i.e., the hardening, the softening and a possible new hardening at about 150~TeV) are clearly visible in the evolution of the spectral index with the energy as shown in Fig.~S10 of the Supplemental Material.\\
\emph{Summary}. The \emph{p}+He spectrum was measured from 46~GeV to 0.46~PeV, using 72 months of data from the DAMPE satellite. The selection of proton+helium, instead of individual proton and helium contributions, effectively prevents cross-contaminations and allows the use of looser selection cuts thus enlarging the statistics and reaching higher energies. The spectrum confirms, with the unprecedented significance of 6.6$\sigma$, the softening feature previously observed in \emph{p} and He individual fluxes. Moreover, the extension at higher energies suggests a new spectral hardening at about 150~TeV. This result also provides an important link between direct and indirect cosmic-ray measurements.\\ 
\emph{Acknowledgments}. The DAMPE mission was funded by the strategic priority science and technology projects in space science of the Chinese Academy of Sciences (CAS). In China, the data analysis was supported by the National Key Research and Development Program of China (No. 2022YFF0503302) and the National Natural Science Foundation of China (Nos. 12220101003, 11921003, 11903084, 12003076 and 12022503), the CAS Project for Young Scientists in Basic Research (No. YSBR061), the Youth Innovation Promotion Association of CAS, the Young Elite Scientists Sponsorship Program by CAST (No. YESS20220197), and the Program for Innovative Talents and Entrepreneur in Jiangsu. In Europe, the activities and data analysis are supported by the Swiss National Science Foundation (SNSF), Switzerland, the National Institute for Nuclear Physics (INFN), Italy, and the European Research Council (ERC) under the European Union’s Horizon 2020 research and innovation programme (No. 851103).
\\
\renewcommand{\thefootnote}{\fnsymbol{footnote}}

\footnotetext[2]{Now at Dipartimento di Matematica e Fisica E. De Giorgi, Università del Salento, I-73100, Lecce, Italy and Istituto Nazionale di Fisica Nucleare (INFN) - Sezione di Lecce, I-73100, Lecce, Italy.}

\footnotetext[3]{Now at Istituto Nazionale Fisica Nucleare (INFN), Sezione di Napoli, IT-80126 Napoli, Italy.}

\footnotetext[4]{Now at Institute of Physics, Ecole Polytechnique F\'{e}d\'{e}rale de Lausanne (EPFL), CH-1015 Lausanne, Switzerland.}

\footnotetext[5]{Now at Department of Nuclear and Particle Physics, University of Geneva, CH-1211 Geneva, Switzerland}

\footnotetext[6]{$ $ Now at Dipartimento di Fisica e Chimica “E. Segrè”, Università degli Studi di Palermo, via delle Scienze ed. 17, I-90128 Palermo, Italy.}

\footnotetext[7]{$ $ Now at Shandong Institute of Advanced Technology (SDIAT), Jinan, Shandong, 250100, China.}

\footnotetext[8]{$ $ Now at Institute of Deep Space Sciences, Deep Space Exploration Laboratory, Hefei 230026, China.}

\footnotetext[5]{$^\P$ Also at School of Physics and Electronic Engineering, Linyi University, Linyi 276000, China.}

\setcounter{figure}{0} 
\setcounter{table}{0} 

\renewcommand{\thefigure}{S\arabic{figure}}
\renewcommand{\thetable}{S\arabic{table}}

\section{Appendix: Supplemental Material}
\label{Appendix}

\section{The DAMPE detector}
\begin{figure}[htbp]
  \centering
  \includegraphics[width=0.45\textwidth]{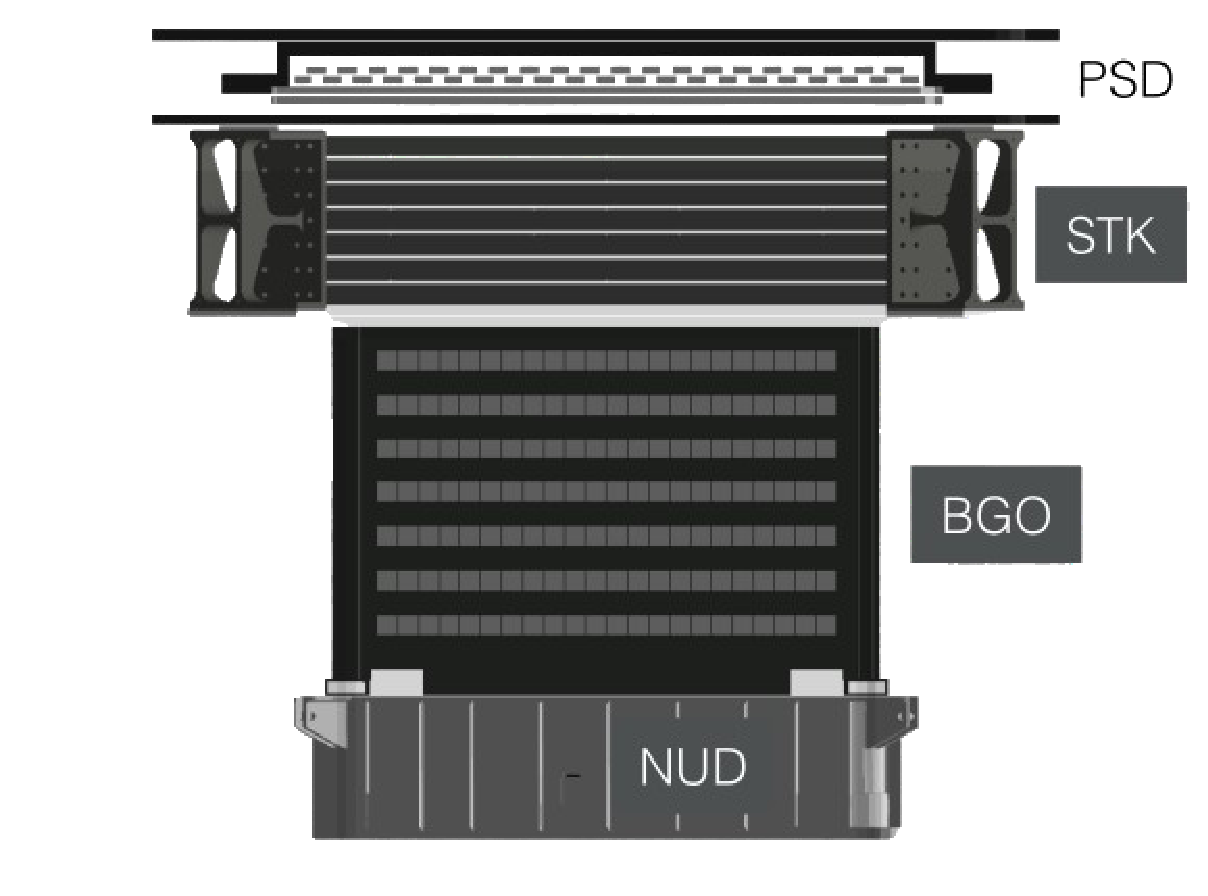} 
  \caption{Schematic view of the DAMPE detector with its subdetectors~\cite{TYKHONOV201843}. See the \emph{Introduction} section of the main text for more details.}%
  \label{fig:DAMPEDet}
\end{figure}

\section{Efficiency validations}
\subsection{HET and track efficiency}
\begin{figure}[htbp]
  \centering
  \includegraphics[width=0.45\textwidth]{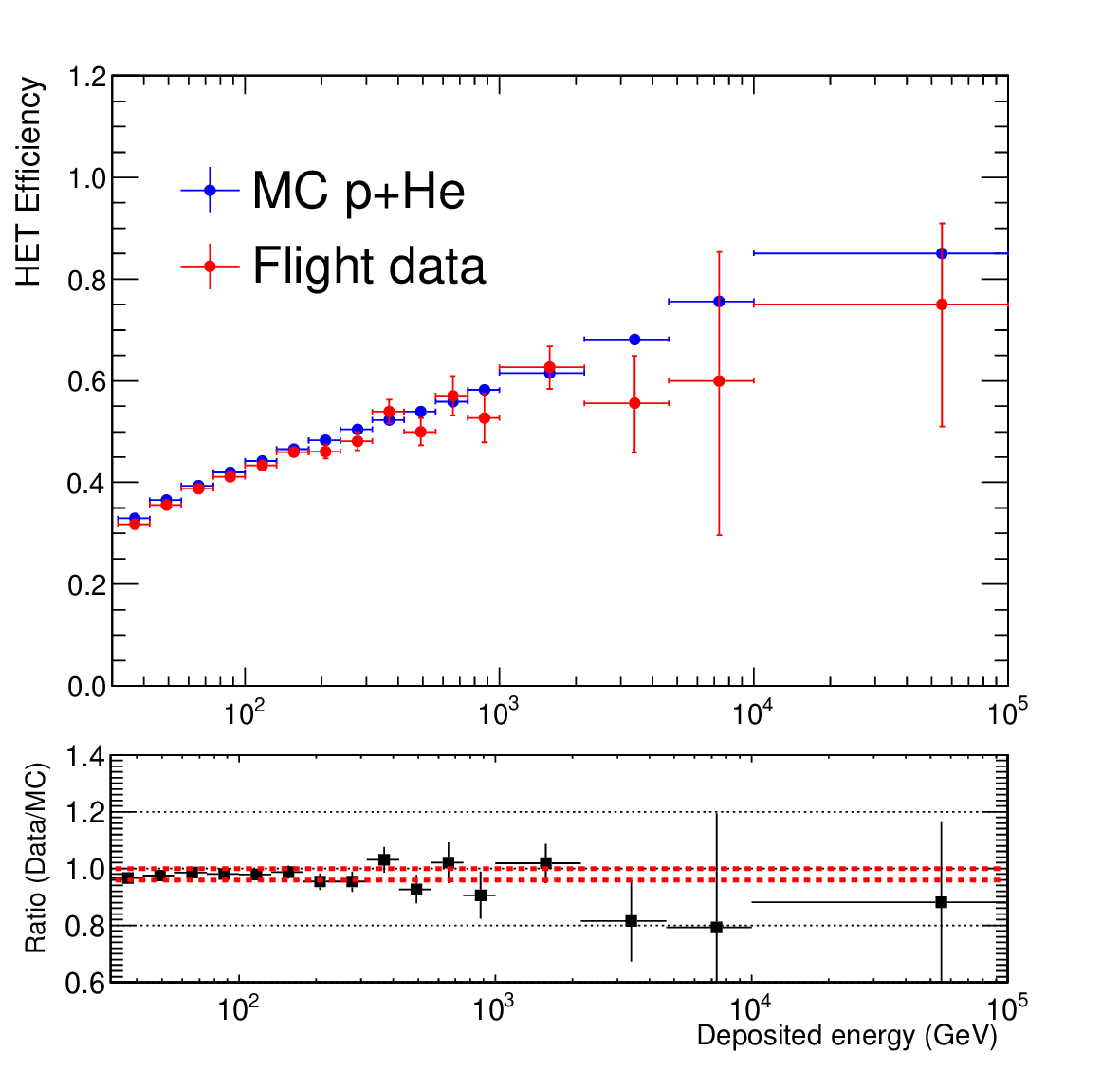} 
  \caption{HET efficiency considering MC (in blue) and flight data (in red) of \emph{p}+He. In the bottom panel, the ratio between MC and flight data is shown and the red dashed lines indicate its variation interval, within the uncertainties}.%
  \label{fig:HET_eff}
\end{figure}

There are four implemented triggers for the DAMPE detector: the Unbiased Trigger (UNBT), the Minimum Ionizing Particle Trigger (MIPT), the Low Energy Trigger (LET) and the High Energy Trigger (HET)~\cite{Zhang_2019}. These triggers are subject to different pre-scaling factors depending on the latitude. The UNBT is the least restrictive and it is used to estimate the HET efficiency, which can be calculated as follows:

\begin{equation}
    \epsilon_{\rm HET} = \frac{N_{\rm HET|\rm UNBT}}{N_{\rm UNBT}},
\end{equation}
where $N_{\rm HET|\rm UNBT}$ is the number of events that pass both the HET and UNBT triggers. Figure~\ref{fig:HET_eff} shows the HET efficiency as a function of the deposited energy in the BGO for MC simulations and flight data. The UNBT sample has a pre-scale factor of 1/512 (1/2048) when the satellite operates in (out of) the geographical latitude range [-20$^o$; 20$^o$]. Therefore, at high energies, the statistical uncertainties of the flight data are expected to be relatively large. 

\begin{figure}[htbp]
  \centering
  \includegraphics[width=0.45\textwidth]{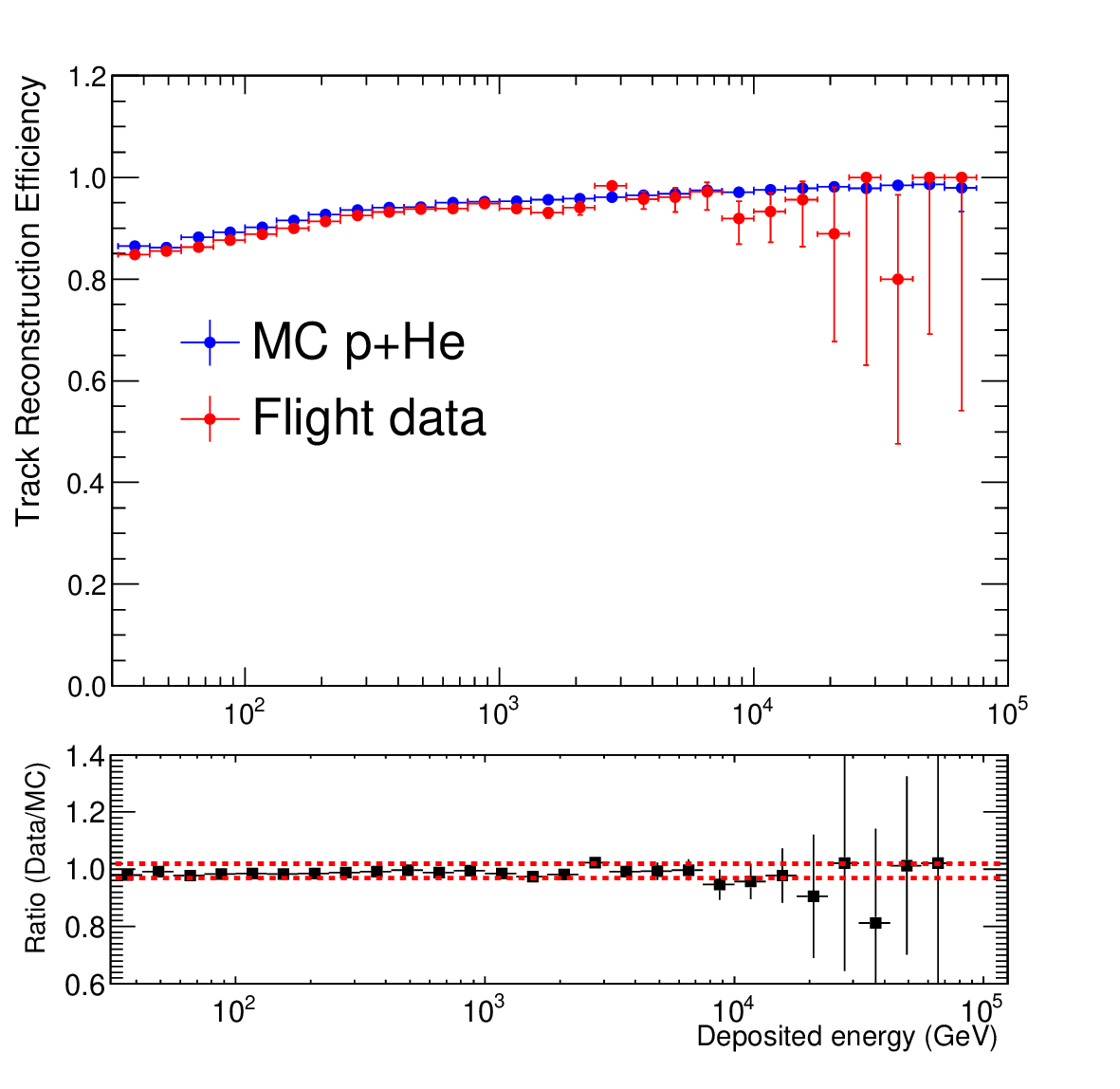}
  \caption{Efficiencies for track selection considering MC (in blue) and flight data (in red) of \emph{p}+He. In the bottom panel, the ratio between MC and flight data is shown and the red dashed lines indicate its variation interval, within the uncertainties}.%
  \label{fig:track_eff}
\end{figure}

Tracks can be determined from hits in the STK and the BGO. However, the former is more precise and is therefore more commonly employed in standard analyses. The STK track efficiency is evaluated by selecting a sample of p and He based on the BGO tracks and PSD charge and considering only the events that pass the STK track selection. The STK track efficiency is given by:
\begin{equation}
    \epsilon_{\rm track} = \frac{N_{\rm {STK|BGO|PSD}}}{N_{\rm{BGO|PSD}}},
\end{equation}
where $N_{\rm{BGO|PSD}}$ is the number of events selected with the BGO track that match the PSD charge, and $N_{\rm{STK|BGO|PSD}}$ is the number of events that pass the STK track selection cuts used in the present analysis. Figure~\ref{fig:track_eff} shows the track selection efficiency as a function of the deposited energy both for MC simulations and flight data.

\subsection{Charge selection efficiency}

\begin{figure*}[htbp]
  \centering
  \subfigure{\label{sub:PSDX}\includegraphics[width=0.45\textwidth]{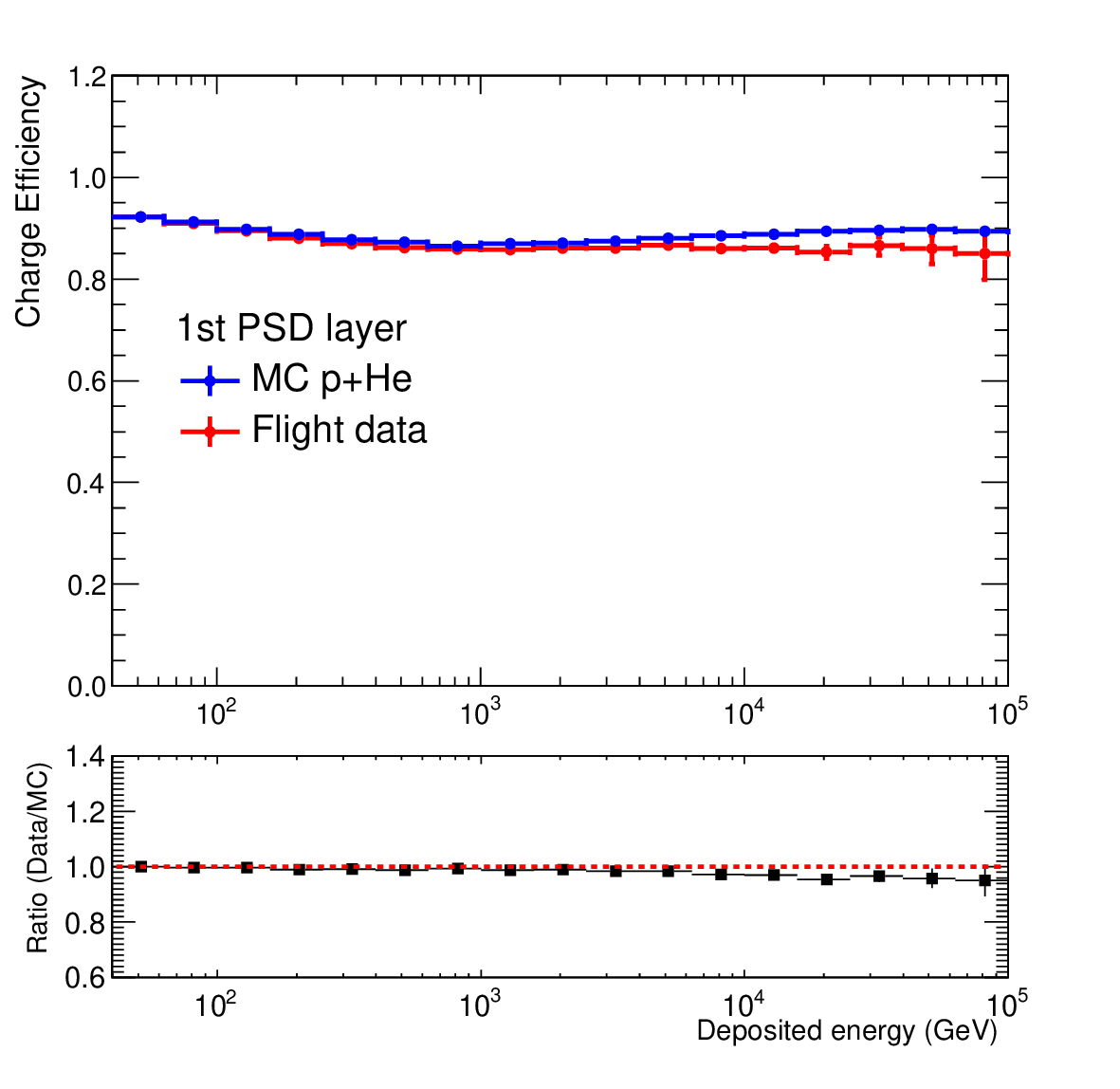}}
  \,
  \subfigure{\label{sub:PSDY}\includegraphics[width=0.45\textwidth]{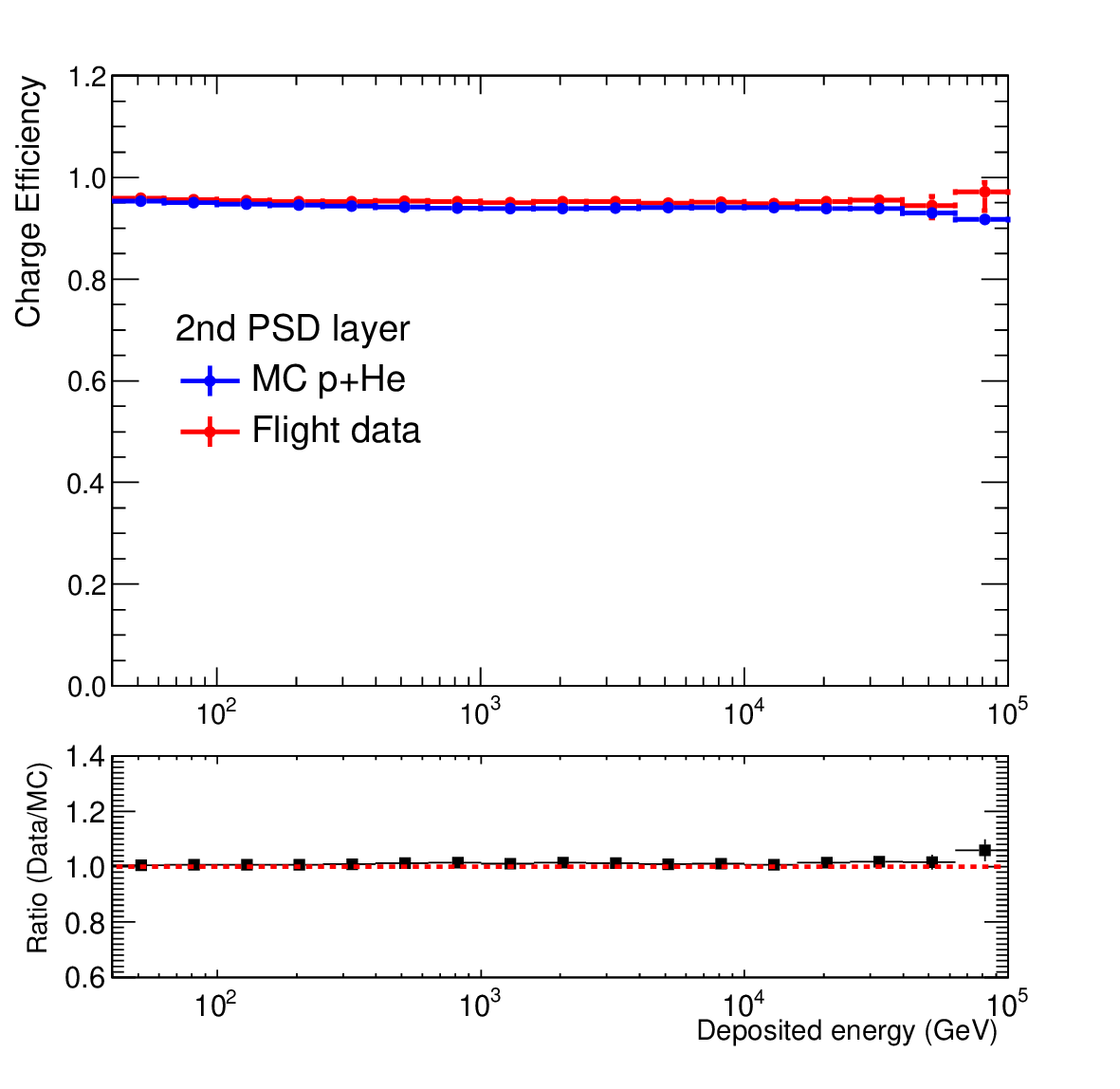}}
  \caption{In the top panels the efficiencies for the first (left) and second (right) layer of the PSD, for MC (in blue) and flight data (in red) of \emph{p}+He are shown. In the bottom panels, the ratio between MC and flight data is presented and the red dashed line indicates its variation within the uncertainties}.%
  \label{fig:charge_eff}
\end{figure*}
Charge selection efficiencies are calculated individually for each PSD layer, using the measurements coming from the first cluster point of the STK track. For example, the efficiency of the first PSD layer is determined by the ratio between the events selected using the charge of both PSD layers and the first cluster point of the STK track ($N_{\rm{PSD_1|PSD_2|STK_1}}$), and the events selected using only the second PSD layer and the first cluster of STK track ($N_{\rm{PSD_2|STK_1}}$):
\begin{equation}
        \epsilon_{\rm{PSD_1}} = \frac{N_{\rm{PSD_1|PSD_2|STK_1}}}{N_{\rm{PSD_2|STK_1}}}.
\end{equation}
An analogous method is used to compute the efficiency of the second layer. Figure~\ref{fig:charge_eff} shows the charge selection efficiencies as a function of the deposited energy for each PSD layer. 

\section{Background contamination}
\begin{figure}[htbp]
     \centering
     \includegraphics[width=0.45\textwidth]{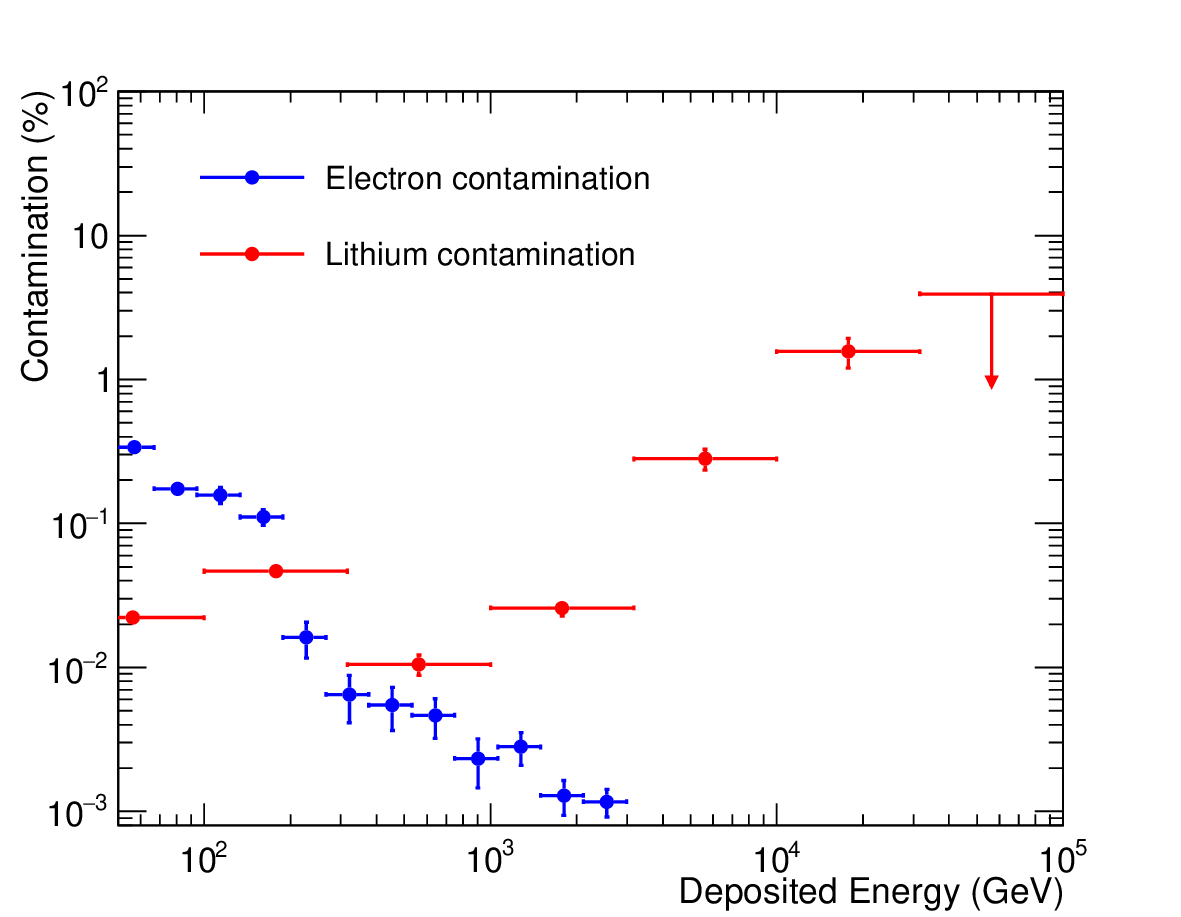}
     \caption{Background in the \emph{p}+He spectrum, from electrons-positrons in blue, and lithium in red.}
     \label{fig:BKG}
 \end{figure}

Combining protons and helium in a single spectrum leaves a very low background, mainly constituted by electrons, positrons and lithium. Electron contamination is calculated by discriminating the morphology of proton-electron showers. The method is thoroughly described in~\cite{electrons}. The lithium background is estimated using the template fit of the energy released in the PSD, based on MC simulations of protons, helium and lithium. Figure~\ref{fig:BKG} shows the estimated electron and lithium backgrounds, which are smaller than 0.4\% up to 10~TeV, and equal to $\sim$1.6\% for energy larger than 10~TeV. 
The possibility of proton and helium interactions outside the considered fiducial volume, but producing events that can survive the \emph{p}+He analysis cuts, has been considered. Dedicated simulations show that this effect can be neglected, being less than 1\% at the highest energies and even smaller at lower energies.

\section{Unfolded flux and systematic uncertainties}
\begin{figure}[htbp]
     \centering
     \includegraphics[width=0.45\textwidth]{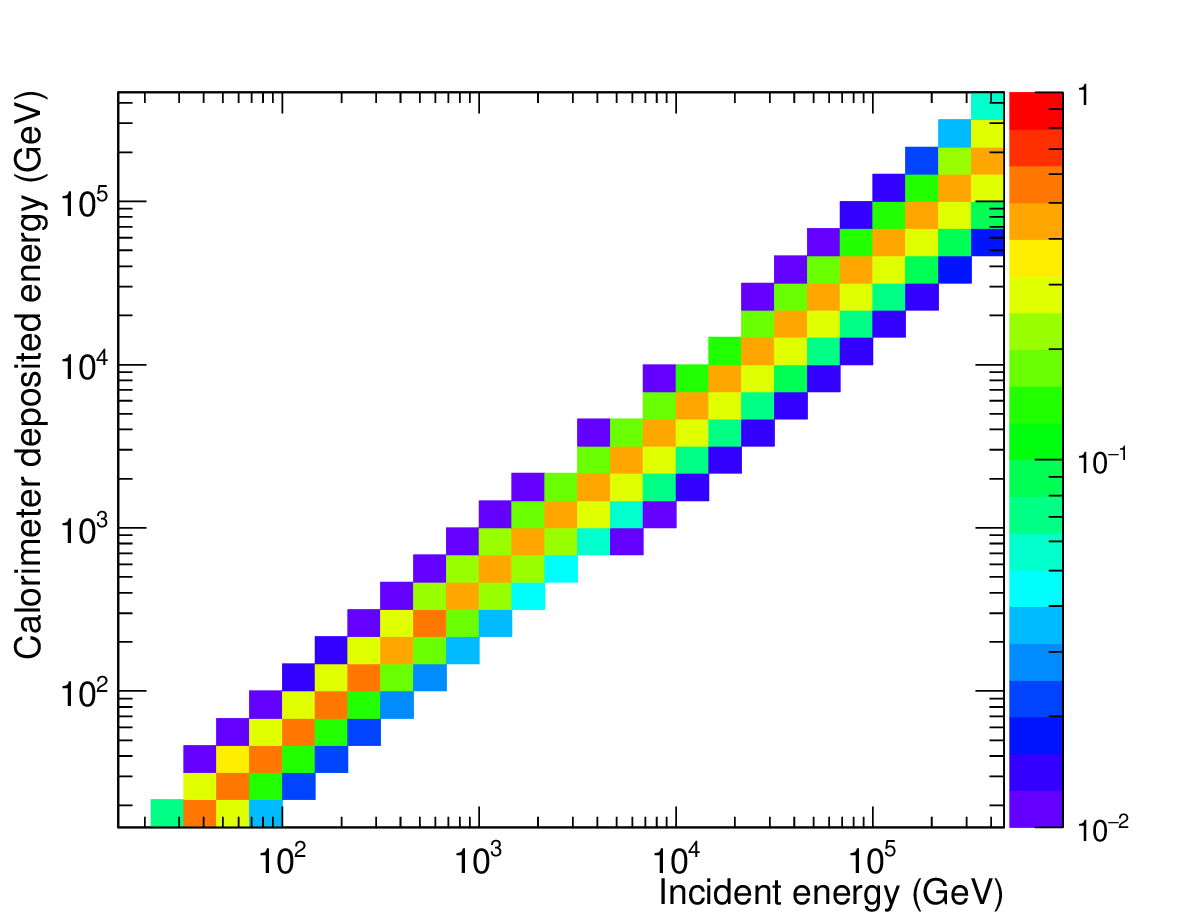}
     \caption{Response matrix derived from MC simulations of p and He after applying the selection cuts. The colors represent the probability that the event in a bin of incident energy migrates to different bins of energy deposited in the BGO calorimeter.}
     \label{fig:response_matrix}
 \end{figure}

The unfolding procedure used to obtain the primary spectrum is described in section \emph{Energy measurement \& unfolding procedure} of the main text. In particular, the number of events in the i-th bin of true energy, $N(E^{\rm i}_{\rm T})$, is derived from the formula (2) of the main text, using the response matrix $P\left(E^{\rm i}_{\rm T} | E^{\rm j}_{\rm O}\right)$ derived from MC. The latter is shown in Figure~\ref{fig:response_matrix}, where the color scale represents the conditional probability that the \emph{p}+He candidates with incident energy $E^{\rm i}_{\rm T}$, are observed with deposited energy $E^{\rm j}_{\rm O}$ in the calorimeter.

\begin{figure}[!ht]
     \centering
     \includegraphics[width=0.45\textwidth]{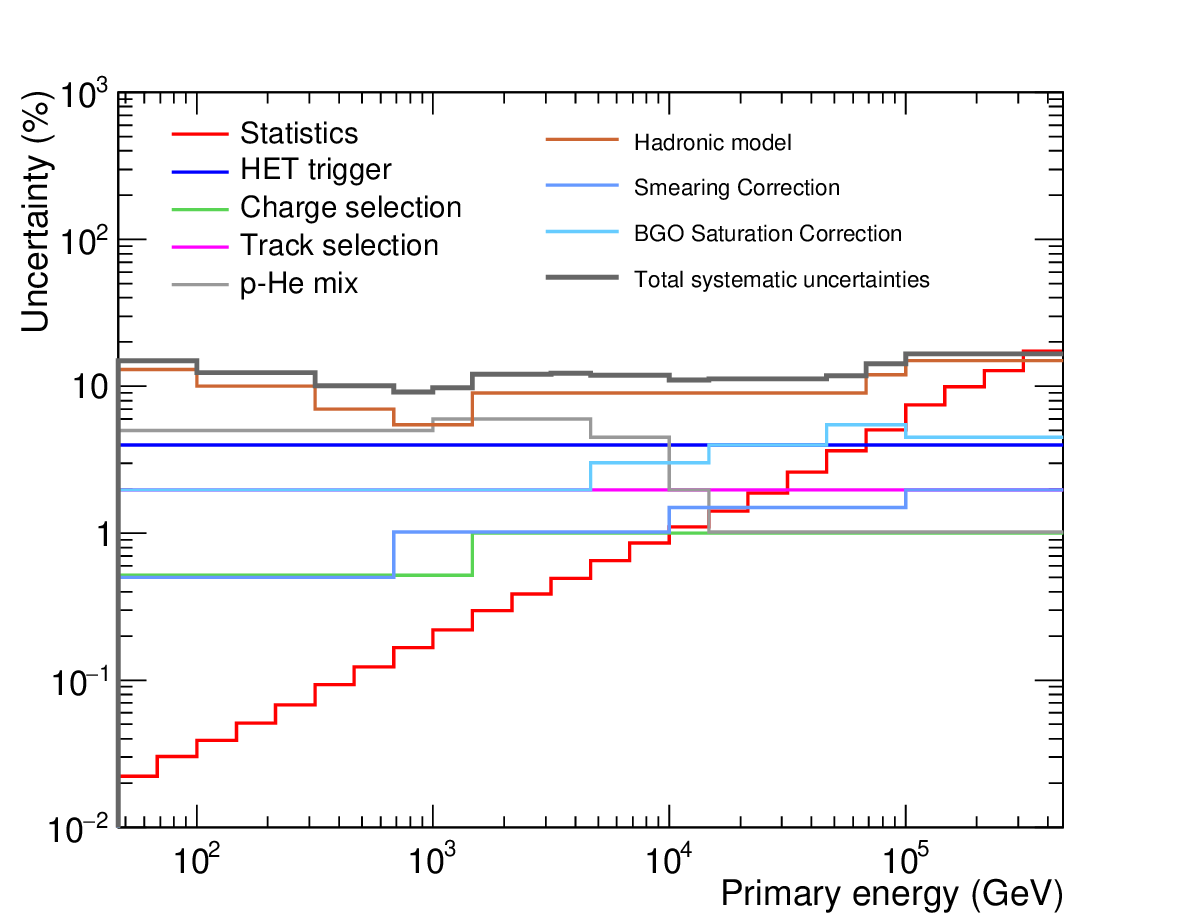}
     \caption{Statistical and systematic uncertainties for the \emph{p}+He spectrum.}
     \label{fig:systematics}
 \end{figure}

Finally, table \ref{Table:SBPL_S} shows the \emph{p}+He flux, (see also Figure~3 of the main text). E, E$_{\rm{low}}$ and E$_{\rm{high}}$ are the median energy and bin edges of the corresponding flux $\Phi$. $\sigma_{\rm{stat}}$ is the statistical error, $\sigma_{\rm{sys}}^{\rm ana}$ and $\sigma_{\rm sys}^{\rm had}$ are the systematic uncertainties on the analysis procedure and on the hadronic model, respectively (see section $Results$ of the main text). Figure~\ref{fig:systematics} shows the statistical error and the relative uncertainty for each considered systematic effect, as described in the paper.

\begin{table*}[htbp]
\centering
\caption{\emph{p}+He flux along with the 1$\sigma$ statistical error and the systematic uncertainties from the analysis and hadronic interaction models. E, E$_{\rm low}$ and E$_{\rm high}$ are the median energy and bin edges of the corresponding flux $\Phi$.}
\setlength{\extrarowheight}{2pt}
\begin{tabular}{ c | c | c | c}
\hline
  E & E$_{\rm low}$ & E$_{\rm high}$ & $\Phi \pm \sigma_{\rm stat}  \pm \sigma_{\rm sys}^{\rm ana} \pm \sigma_{\rm sys}^{\rm had}$ \\
  
  [GeV] & [GeV] & [GeV] & [GeV$^{-1}m^{-2}s^{-1}sr^{-1}$]\\
\hline
5.62 $\times$10$^{1}$ & 4.64 $\times$10$^{1}$ & 6.81 $\times$10$^{1}$ & (3.31 $\pm$ 0.00073 $\pm$ 0.24  $\pm$ 0.49) $\times$10$^{-1}$ \\
8.25 $\times$10$^{1}$ & 6.81 $\times$10$^{1}$ & 1.00 $\times$10$^{2}$ & (1.17 $\pm$ 0.00035 $\pm$ 0.08  $\pm$ 0.17) $\times$10$^{-1}$ \\
1.21 $\times$10$^{2}$ & 1.00 $\times$10$^{2}$ & 1.47 $\times$10$^{2}$ & (4.12 $\pm$ 0.0016 $\pm$ 0.30  $\pm$ 0.51)  $\times$10$^{-2}$ \\
1.78 $\times$10$^{2}$ & 1.47 $\times$10$^{2}$ & 2.15 $\times$10$^{2}$ & (1.45 $\pm$ 0.0074 $\pm$ 0.11  $\pm$ 0.18)  $\times$10$^{-2}$ \\
2.61 $\times$10$^{2}$ & 2.15 $\times$10$^{2}$ & 3.16 $\times$10$^{2}$ & (5.13 $\pm$ 0.0035 $\pm$ 0.37  $\pm$ 0.63)  $\times$10$^{-3}$ \\
3.83 $\times$10$^{2}$ & 3.16 $\times$10$^{2}$ & 4.64 $\times$10$^{2}$ & (1.81 $\pm$ 0.0017 $\pm$ 0.13  $\pm$ 0.18)  $\times$10$^{-3}$ \\
5.62 $\times$10$^{2}$ & 4.64 $\times$10$^{2}$ & 6.81 $\times$10$^{2}$ & (6.46 $\pm$ 0.0079 $\pm$ 0.47  $\pm$ 0.65)  $\times$10$^{-4}$ \\
8.25 $\times$10$^{2}$ & 6.81 $\times$10$^{2}$ & 1.00 $\times$10$^{3}$ & (2.34 $\pm$ 0.0039 $\pm$ 0.17  $\pm$ 0.22)  $\times$10$^{-4}$ \\
1.21 $\times$10$^{3}$ & 1.00 $\times$10$^{3}$ & 1.47 $\times$10$^{3}$ & (8.60 $\pm$ 0.019 $\pm$ 0.69  $\pm$ 0.84)   $\times$10$^{-5}$ \\
1.78 $\times$10$^{3}$ & 1.47 $\times$10$^{3}$ & 2.15 $\times$10$^{3}$ & (3.17 $\pm$ 0.0094 $\pm$ 0.26  $\pm$ 0.38)  $\times$10$^{-5}$ \\
2.61 $\times$10$^{3}$ & 2.15 $\times$10$^{3}$ & 3.16 $\times$10$^{3}$ & (1.20 $\pm$ 0.0046 $\pm$ 0.10  $\pm$ 0.15)  $\times$10$^{-5}$ \\
3.83 $\times$10$^{3}$ & 3.16 $\times$10$^{3}$ & 4.64 $\times$10$^{3}$ & (4.63 $\pm$ 0.023 $\pm$ 0.39  $\pm$ 0.57)  $\times$10$^{-6}$ \\
5.62 $\times$10$^{3}$ & 4.64 $\times$10$^{3}$ & 6.81 $\times$10$^{3}$ & (1.76 $\pm$ 0.012 $\pm$ 0.14  $\pm$ 0.21)  $\times$10$^{-6}$ \\
8.25 $\times$10$^{3}$ & 6.81 $\times$10$^{3}$ & 1.00 $\times$10$^{4}$ & (6.65 $\pm$ 0.057 $\pm$ 0.52  $\pm$ 0.79)  $\times$10$^{-7}$ \\
1.21 $\times$10$^{4}$ & 1.00 $\times$10$^{4}$ & 1.47 $\times$10$^{4}$ & (2.53 $\pm$ 0.028 $\pm$ 0.16  $\pm$ 0.28)  $\times$10$^{-7}$ \\
1.78 $\times$10$^{4}$ & 1.47 $\times$10$^{4}$ & 2.15 $\times$10$^{4}$ & (0.96 $\pm$ 0.014 $\pm$ 0.06  $\pm$ 0.11)  $\times$10$^{-7}$ \\
2.61 $\times$10$^{4}$ & 2.15 $\times$10$^{4}$ & 3.16 $\times$10$^{4}$ & (3.57 $\pm$ 0.067 $\pm$ 0.24  $\pm$ 0.40)  $\times$10$^{-8}$ \\
3.83 $\times$10$^{4}$ & 3.16 $\times$10$^{4}$ & 4.64 $\times$10$^{4}$ & (1.25 $\pm$ 0.033 $\pm$ 0.08  $\pm$ 0.14)  $\times$10$^{-8}$ \\
5.62 $\times$10$^{4}$ & 4.64 $\times$10$^{4}$ & 6.81 $\times$10$^{4}$ & (4.04 $\pm$ 0.15 $\pm$ 0.31  $\pm$ 0.48)   $\times$10$^{-9}$ \\
8.25 $\times$10$^{4}$ & 6.81 $\times$10$^{4}$ & 1.00 $\times$10$^{5}$ & (1.35 $\pm$ 0.068 $\pm$ 0.10 $\pm$ 0.19)   $\times$10$^{-9}$ \\
1.21 $\times$10$^{5}$ & 1.00 $\times$10$^{5}$ & 1.47 $\times$10$^{5}$ & (4.49 $\pm$ 0.33 $\pm$ 0.32 $\pm$ 0.74)    $\times$10$^{-10}$ \\
1.78 $\times$10$^{5}$ & 1.47 $\times$10$^{5}$ & 2.15 $\times$10$^{5}$ & (1.59 $\pm$ 0.16 $\pm$ 0.11 $\pm$ 0.26)    $\times$10$^{-10}$ \\
2.61 $\times$10$^{5}$ & 2.15 $\times$10$^{5}$ & 3.16 $\times$10$^{5}$ & (5.70 $\pm$ 0.73 $\pm$ 0.40 $\pm$ 0.95)    $\times$10$^{-11}$ \\
3.83 $\times$10$^{5}$ & 3.16 $\times$10$^{5}$ & 4.64 $\times$10$^{5}$ & (2.07 $\pm$ 0.36 $\pm$ 0.15 $\pm$ 0.34)    $\times$10$^{-11}$ \\
\hline
\end{tabular}
\label{Table:SBPL_S}
\end{table*}

\section{Spectral features}
\begin{figure}[htbp] 
     \centering
     \includegraphics[width=0.45\textwidth]{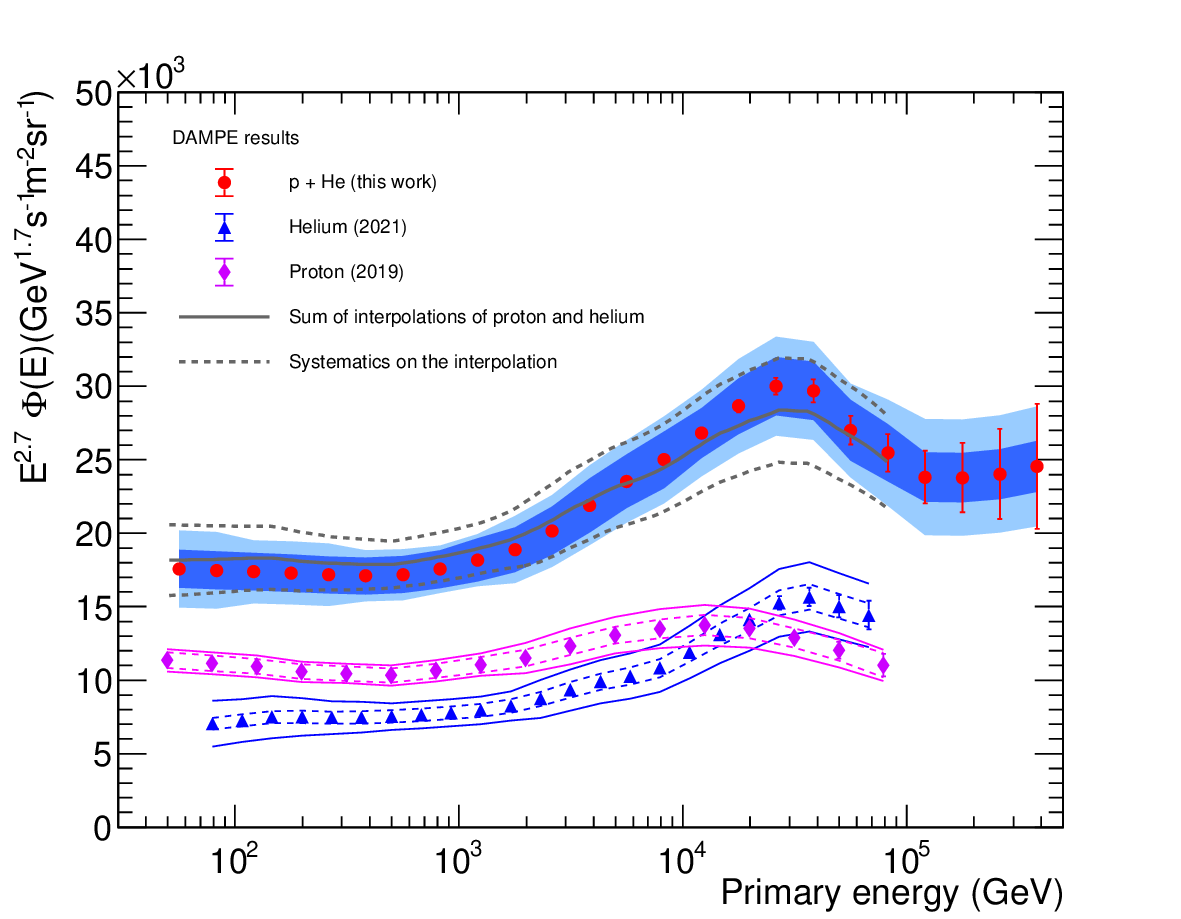}
     \caption{Comparison among proton~\cite{doi:10.1126/sciadv.aax3793} (magenta diamonds), helium~\cite{PhysRevLett.126.201102} (blue triangles) and \emph{p}+He (this work, red circles) spectra from DAMPE. The systematic uncertainties of the \emph{p}+He spectrum are shown with blue and light-blue bands, while the proton and helium individual spectra systematic uncertainties are indicated with dashed and continuous lines. The total systematic uncertainties are denoted by the outer band (continuous line), while the inner band (dashed line) doesn’t include the uncertainties related to the hadronic model. The error bars correspond to the 1$\sigma$ statistical uncertainties. Finally, the sum of the interpolated proton and helium individual spectra, along with the sum of the respective systematics, is depicted by a continuous grey line and dashed grey line respectively, superimposed on the \emph{p}+He spectrum.}
     \label{fig:DAMPEComp}
 \end{figure}

The result from this work was compared with the DAMPE proton~\cite{doi:10.1126/sciadv.aax3793} and helium~\cite{PhysRevLett.126.201102} individual spectra, as can be seen in Figure~\ref{fig:DAMPEComp}, which shows good internal consistency.

\begin{figure}[htbp] 
     \centering
     \includegraphics[width=0.45\textwidth]{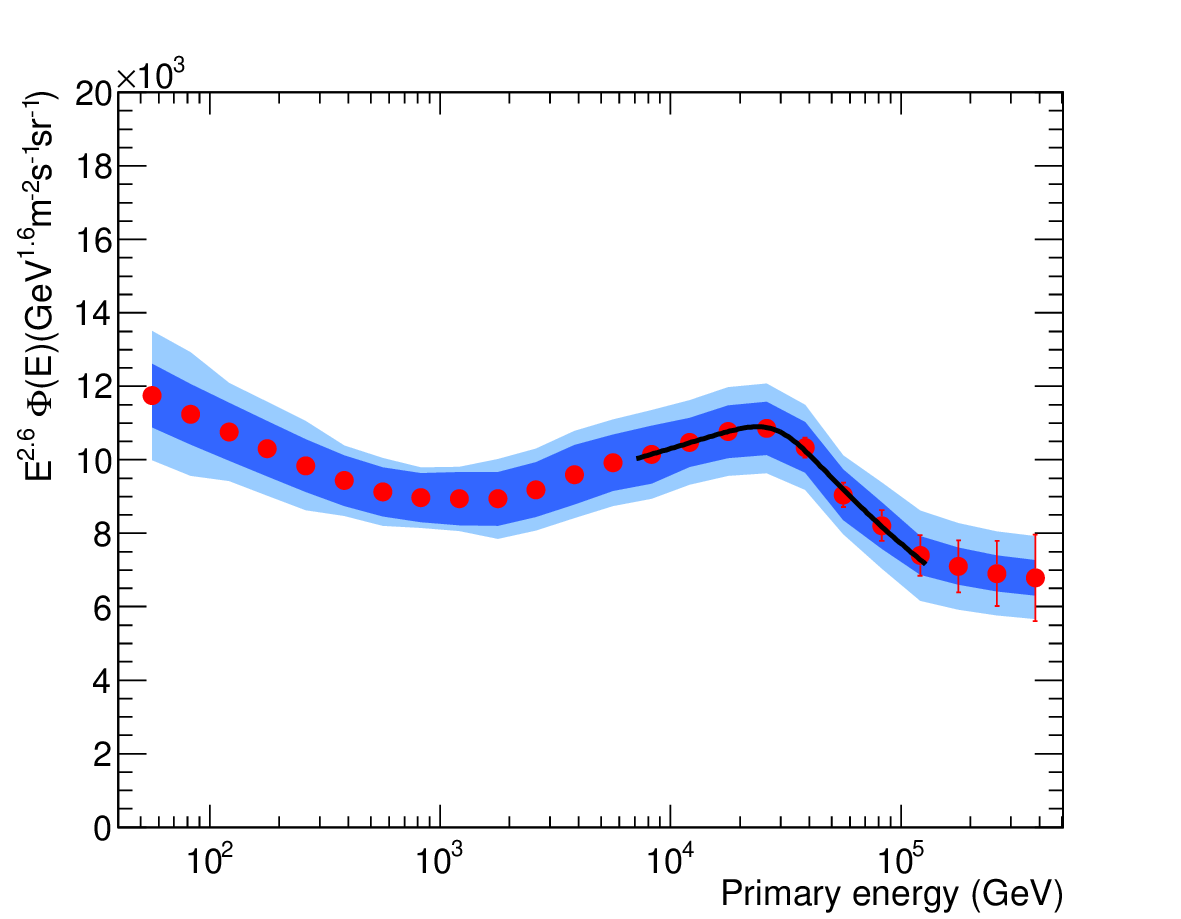}
     \caption{Fit of the \emph{p}+He spectrum (in red) with a SBPL function (in black). Statistical uncertainties are represented by error bars. In shaded bands, the systematic uncertainties on the analysis (inner band) and total (outer band) are shown.}
     \label{fig:Fit}
 \end{figure}

Moreover, the proton+helium spectrum exhibits a softening feature that is well described by a smoothly broken power law:
\begin{equation}
\Phi(E)=\Phi_{0}\left(\frac{E}{\mathrm{TeV}}\right)^{-\gamma}\left[1+\left(\frac{E}{E_{\rm B}}\right)^{s}\right]^{\Delta \gamma / s},
\label{fitfunc}
\end{equation}
where $\Phi_{0}$ is the flux normalization, $\gamma$ the spectral index before the break energy ($E_{\rm B}$), $\Delta \gamma$ the difference between the spectral index before and after the break, and \emph{s} the smoothness of the break. To properly account for the systematic uncertainties, a set of independent nuisance parameters are adopted and multiplied to the input model, using a similar procedure as the one used in~\cite{PhysRevD.95.082007,electrons, doi:10.1126/sciadv.aax3793, PhysRevLett.126.201102}. The $\chi^2$ function is defined as follows:

{\footnotesize
\begin{align}
    \chi^2 = &\sum_{i=k}^{n}\sum_{j=k}^{n}[ \Phi\left(E_{\rm i}\right)S\left(E_{\rm i}; \rm w\right)-\Phi_{\rm i}] C_{i,j}^{-1} [ \Phi\left(E_{\rm j}\right)S\left(E_{\rm j}; \rm w\right)-\Phi_{\rm j}] \nonumber \\&+ \sum_{l=1}^{m}\left( \frac{1-w_{\rm l}}{\tilde{\sigma}_{\rm {sys,l}}}\right)^2,
\end{align}
}

where $E_{\rm i}$ is the median energy, $\Phi_{\rm i}$ the flux in the i-th energy bin, $\Phi(E_{\rm i})$ is the model predicted flux in each corresponding energy bin, $S(E_{\rm i}, \rm w)$ is a piece-wise function defined by its value $w_{\rm j}$ in the corresponding energy range covered by the j-th nuisance parameter, $C_{i,j}$ represents the covariance matrix of the fluxes obtained from toy MC simulations when estimating the statistical uncertainties and $\tilde{\sigma}_{\rm {sys,j}}=(\sigma_{\rm sys}^{\rm ana})/\Phi$ is the relative systematic uncertainty of the data in such an energy range. The fit is performed in the energy range from 7 to 130~TeV, using 2 nuisance parameters. In order to account for the uncertainties on the fit parameters resulting from the selected hadronic model, the same fit is performed on the spectrum computed with MC samples simulated using the FLUKA DPMJET-3 and GEANT4-QGSP\_BERT models. The difference between the two fits is taken as a second uncertainty on the parameters. The results of the fit are $\Phi_{0}$ = (1.35~$\pm$~0.09)~$\times$~10$^{-4}$~GeV$^{-1}$m$^{-2}$s$^{-1}$sr$^{-1}$, $\gamma$ = $2.512_{-0.024-0.00}^{+0.021+0.01}$, $\Delta\gamma$ = $-0.427_{-0.057-0.00}^{+0.066+0.066}$, E$\mathrm{_{B}}$ = $28.8_{-4.4-0.0}^{+6.2+2.9}$~TeV, and $ \chi^{2}$/dof = 0.9/2. The parameter s was fixed to 5, which is consistent with the DAMPE proton and helium fit of the softening~\cite{doi:10.1126/sciadv.aax3793, PhysRevLett.126.201102}. To estimate the significance of the softening, the same energy region has been fitted with a single power-law function, giving a result of $ \chi^{2}$/dof = 48.14/4, which translates to a significance of 6.6$\sigma$. The SBPL fit of the softening is shown in Figure~\ref{fig:Fit}.
 \begin{figure}[htbp] 
     \centering
     \includegraphics[width=0.45\textwidth]{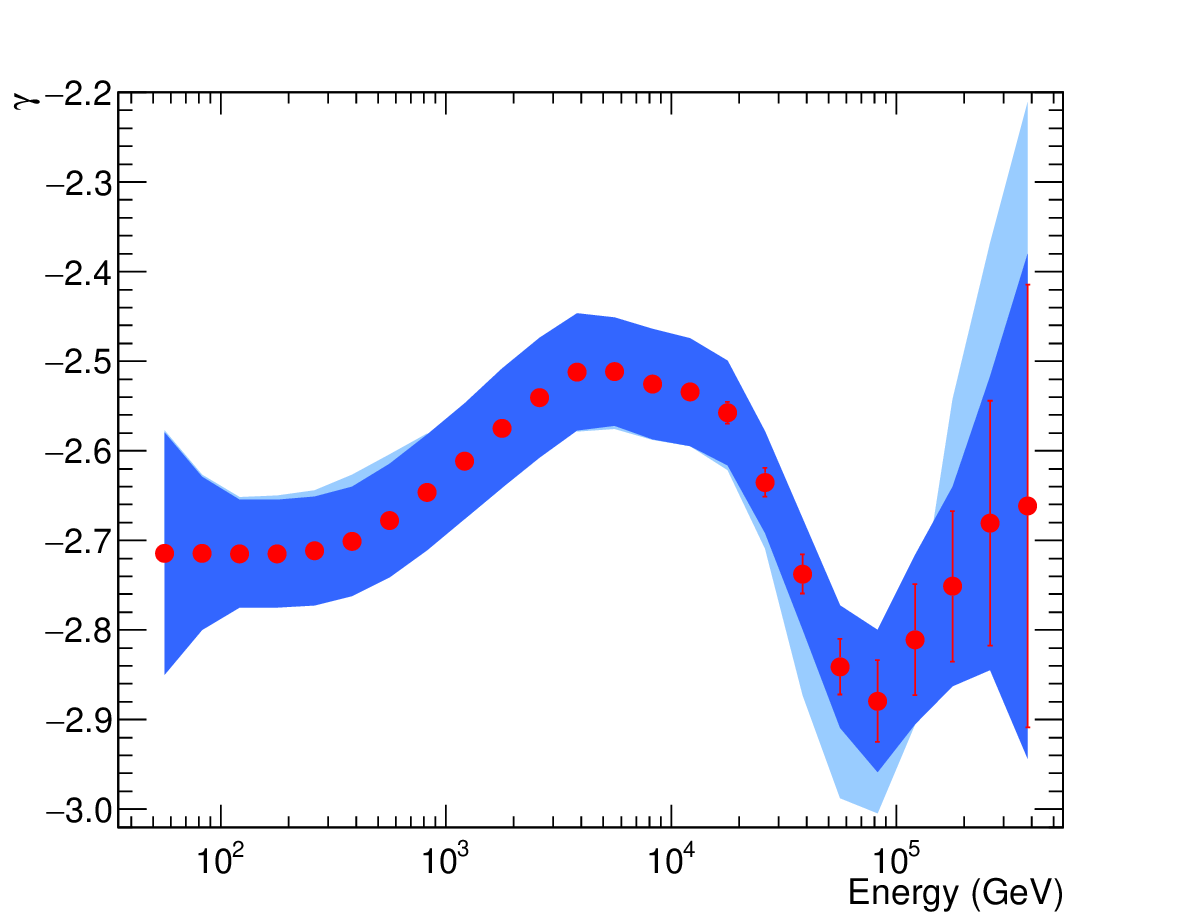}
     \caption{Variation of the spectral index with the energy for DAMPE \emph{p}+He data. The spectral index results from the fit of the \emph{p}+He spectrum in sliding energy windows of $\pm$2 bins, considering smaller intervals in the first and last two bins. The uncertainties are indicated with error bars for statistics, a blue band for systematics on the analysis, and a light blue band for systematics on the hadronic model (see text for additional details).}
     \label{fig:Fitgamma}
 \end{figure}

In order to further study the evolution of the spectral index ($\gamma$) with energy, the \emph{p}+He spectrum has been fitted using sliding windows of $\pm$2 bins (while smaller energy intervals are used for the first and last two points). The result is shown in Figure~\ref{fig:Fitgamma}, where the data points represent the $\gamma$ value with its statistical uncertainty and the blue inner band indicates the quadratic sum of statistical and systematic uncertainties on the analysis. The light blue additional band includes also the systematics on the hadronic model, computed as done for the fit of the softening. Specifically, the sliding window technique has been applied to the energy spectrum obtained by assuming either the default or the alternative hadronic model (see main text [link]). The difference between the $\gamma$ values obtained in the two cases is taken as the estimate of the uncertainty on the hadronic model. This is summed in quadrature with the uncertainty shown by the blue inner band to give the total uncertainty. The spectral hardening and softening can be easily recognized in the figure. In addition, the gamma evolution with the energy does confirm the suggestion of a new hardening at about 150~TeV. A spectral fit with a broken power law was performed above 30~TeV disfavouring at 1.5$\sigma$ a single power law hypothesis.

\bibliographystyle{apsrev4-2}
\bibliography{CR}
\end{document}